%% file: main_arxiv.tex
\documentclass{article}
\usepackage{PRIMEarxiv}
\usepackage[utf8]{inputenc}
\usepackage[T1]{fontenc}
\usepackage{microtype}
\usepackage{amsmath,amssymb,graphicx,booktabs}
\usepackage[section]{placeins}
\usepackage[hidelinks]{hyperref}
\usepackage{url}
\usepackage{xcolor}
\usepackage[numbers,sort&compress]{natbib}

\newcommand{\qj}{q_j}
\newcommand{\Dj}{\Delta_j}
\newcommand{\EN}{E(N)}
\newcommand{\taueff}{\tau_{\mathrm{eff}}}
\newcommand{\Nc}{N_c}

\title{Predicting the scale limits of social mechanisms in agent societies}
\author{
  Zengqing Wu \\
  University of Osaka \\
  Osaka, Japan \\
  \And
  Chuan Xiao\thanks{Corresponding author: \texttt{chuanx@ist.osaka-u.ac.jp}} \\
  University of Osaka \\
  Osaka, Japan \\
}
\date{}

\begin{document}
\maketitle

\begin{abstract}
\input{sections_nc/abstract}
\end{abstract}

\input{sections_nc/introduction}

\section{Results}

\input{sections_nc/results_audit}
\input{sections_nc/results_single_terms}
\input{sections_nc/results_product_law}
\input{sections_nc/results_formats}
\input{sections_nc/results_directions}
\input{sections_nc/results_prospective}

\input{sections_nc/discussion}

\input{sections_nc/methods}

\input{sections_nc/availability}

\input{sections_nc/backmatter}

\bibliographystyle{unsrtnat}
\bibliography{references}

\clearpage
\appendix
\renewcommand{\thesection}{S\arabic{section}}
\renewcommand{\thetable}{S\arabic{table}}
\renewcommand{\thefigure}{S\arabic{figure}}
\renewcommand{\theHsection}{S\arabic{section}}
\renewcommand{\theHtable}{S\arabic{table}}
\renewcommand{\theHfigure}{S\arabic{figure}}
\setcounter{section}{0}
\setcounter{table}{0}
\setcounter{figure}{0}

\begin{center}
{\LARGE\bfseries Supplementary Information}
\end{center}

\input{si_body}

\end{document}

%% file: sections_nc/abstract.tex
Societies of interacting language-model agents offer a controllable and
repeatable way to study collective behaviour at scales that would be
difficult to test with people. Their scientific value, however, depends on
whether a social mechanism that works in a small group still operates when
thousands of agents interact, and testing this directly requires costly
large-scale runs. Here we introduce an audit that predicts a
mechanism's fate as a population grows. It asks how often the mechanism can
act, whether agents use the information it supplies, and whether the
measurement itself creates apparent scale effects. Controlled experiments
show that a single structural term can decide whether reciprocity,
consensus or punishment survives scaling. For gossip, the population at
which the mechanism fails is set by the reach and lifetime of its
messages. In
language-model societies, agents respond not only to social information
but to how it is expressed: counts and percentages led to different
scale behaviour. Predictions made before execution held on third-party
code and a second model family, while a failed prediction exposed
the boundary of the finding. The audit provides a prospective way to
decide which social mechanisms can be interpreted across population
scales.

%% file: sections_nc/introduction.tex
\section{Introduction}

Societies built from language-model agents offer a new experimental
window on collective behaviour. Unlike traditional rule-based agents,
language models can interpret open-ended descriptions, occupy different
social roles and respond to the history of an interaction. Because the
same population can be copied, controlled and rerun, these systems make it
possible to explore social processes and counterfactual worlds that would
be costly, slow or impractical to study with human participants. This
possibility has prompted calls to study agents as behavioural entities
that can be observed and experimentally perturbed, rather than only as
models to be benchmarked \cite{chen2026ai}. The field has moved quickly.
Early generative-agent work showed that a small town of language-model
agents could produce believable daily social life \cite{park2023generative},
and generative agent-based modelling frameworks connected such agents to
the questions of classical social simulation \cite{vezhnevets2023generative}.
Individual language models can predict aggregate outcomes of survey
experiments \cite{ashokkumar2026large}, play repeated social dilemmas
\cite{akata2025playing}, reproduce patterns of trust between partners
\cite{sakamoto2025value}, and generate recognizable features of citation
networks \cite{ji2026leveraging}. General-purpose platforms now populate
simulations of social media, cities and markets
\cite{tang-etal-2025-gensim, gao2024large, agentsociety2025,
yang2024oasis, al2024project}, in some cases assembling $10^{4}$ to
$10^{6}$ interacting agents. The community is meanwhile debating what
such simulations can support, including where their appropriate boundary
lies \cite{wu2026position}, how a persistent validation gap should be
closed \cite{touzel2026position}, and how often current practice violates
basic validity principles, with reported collective phenomena vanishing
or reversing once those principles are enforced \cite{zhou2025pimmur}. At this point, the validity of an
artificial society depends on more than whether one agent resembles one
person: it also depends on whether the processes connecting agents remain
meaningful as the population grows.

Scaling up is not a neutral change. In a small group, two individuals may
meet repeatedly. In a large population, the same pair may never meet
again. A piece of gossip sent to four people can cover most of a group of
five but almost none of a group of five thousand. A public record can
remove that bottleneck by making the same event visible to everyone.
Language-model agents introduce a further complication because the same
social fact can be expressed as a count, a percentage or a story about a
named person, and the model need not treat these forms as equivalent.
The realism of natural-language interaction can therefore work against
explanation when it obscures the link between local behaviour and
system-level emergence \cite{zeng2026too}. Consequently, a change in the
output of a large simulation can have at least three origins: the social
mechanism lost structural reach, the agents did not use the information
supplied to them, or the observation protocol manufactured an apparent
change. An aggregate curve alone cannot separate these explanations
\cite{watts2014common, hofman2017prediction}. The ambiguity is expensive,
since each large run of a language-model society consumes substantial
computation, and resolving it by trial and error means paying that cost
repeatedly. A prediction that precedes the run turns the question into a
check.

Decades of work show that cooperation in large populations can be
sustained when information and institutions compensate for anonymity
\cite{kandori1992social, milgrom1990role, acemoglu2020sustaining}.
Experiments likewise show that group size often acts through monitoring,
observability and critical mass rather than as a cause on its own
\cite{wu2020cooperation, pereda2019large, yoeli2013powering,
centola2018experimental}. Recent studies have begun to vary population
size in language-model groups and to compare their collective dynamics
with those of rule-based particles. They reveal changes in convention
formation, consensus, segregation and collective performance
\cite{flint2025group, bertalanivc2026ringelmann, tanaka2026collective,
zomer2026unraveling}. Related theory shows why maintaining stable
cooperation requires proportionally more rounds of gossip as the population grows
\cite{kawakatsu2024mechanistic}. Recent language-model studies further
show that simulation results can be sensitive to seemingly minor design
choices \cite{ye2026stop}, while natural-language gossip can sustain
cooperation in small groups of five or nine agents \cite{zhu2026talk}. Increasing
the size of the language model may improve simulation fidelity
\cite{ziems2026will}, but that is distinct from increasing the number of
agents and the interactions among them. These studies establish that scale
and interaction structure matter. What is still missing is a procedure
that can predict, before a large run, which part of a mechanism will
survive, which part will disappear and where the transition should occur.

We address this gap by treating scale behaviour as a property of three
objects that can be examined separately: the mechanism, the agents and the
observation plan. For each route by which a mechanism can change
behaviour, the audit asks three plain questions. How frequently does this
route reach a decision? Does the agent respond when it does? And can the
response be compared fairly across population sizes? Structural terms can
often be read from code, whereas responses of language-model agents must
be measured directly. The resulting path-level predictions are combined
only when all active routes are known; otherwise the aggregate outcome is
declared unidentified rather than guessed.

The experiments reveal three general findings. First, the scale fate of a
social mechanism can hinge on a single structural quantity, and the
failure point of gossip follows a simple law set by message reach and
lifetime. Second, language-model agents respond to the representation of
social information: counts and explicit percentages can place an
otherwise identical mechanism in different scale regimes. Third,
directional responses are more reproducible across model versions than
their absolute levels. Predictions made before execution were tested on
unmodified external code and across model families. Successes
establish the reach of the audit, while a failed prediction narrows its
scope. Together, the results ask whether the fate of a mechanism at scale
can be computed from parts that are measurable before the full society is
run.

%% file: sections_nc/results_audit.tex
\subsection{An audit from mechanism to population}

The audit rests on a simple accounting idea. A social mechanism changes
behaviour through one or more information pathways. The contribution of
each pathway depends on how often it reaches a decision and how strongly
an agent responds when it does. The first quantity can often be computed
from the implementation before an experiment runs, and the second can be
measured one decision at a time. We write the total effect as
\begin{equation}
\EN \;=\; \textstyle\sum_j \qj(N)\,\Dj(N) \;+\; \text{interactions},
\label{eq:aggregate}
\end{equation}
where $\EN$ is the mechanism's effect relative to a matched baseline
population without it, $\qj(N)$ is the exposure or coverage of pathway
$j$, and $\Dj(N)$ is the behavioural change produced by one exposure. For gossip, a pathway is
the route by which a report about the current partner reaches a decision:
$\qj(N)$ is the fraction of relevant decisions at which the report is
visible, whereas $\Dj(N)$ is the change in giving conditional on seeing
it. Their product is the expected contribution of that pathway to the
aggregate effect. Equation~\eqref{eq:aggregate} is a
path-wise decomposition, conceptually related to, but not an estimator of,
causal mediation effects \cite{pearl2022direct,
vanderweele2015explanation}.

\begin{figure}[tbp]\centering
\includegraphics[width=\linewidth]{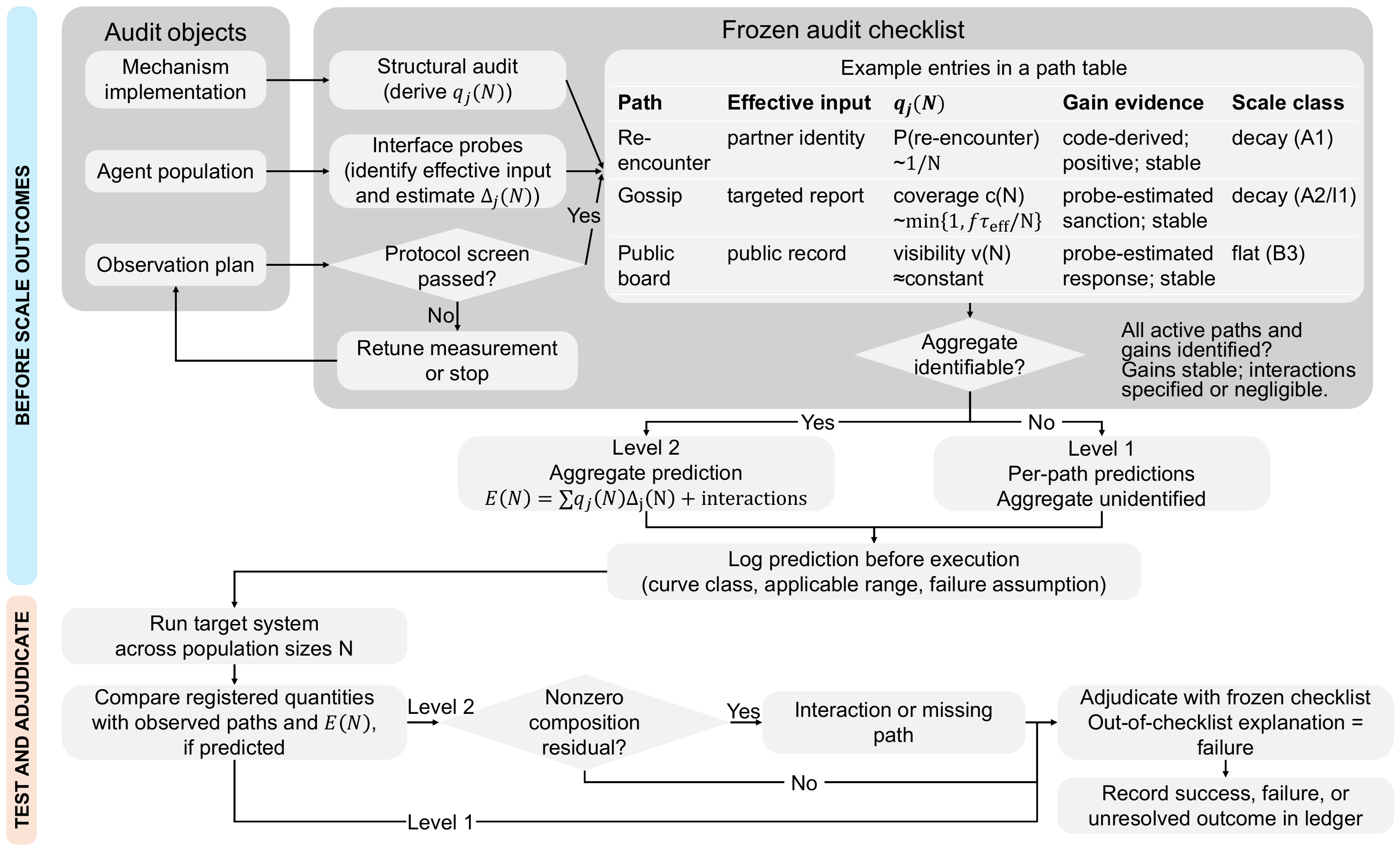}
\caption{Prospective audit and adjudication workflow. Before any scale
outcome exists, the mechanism implementation yields structural quantities
$q_j(N)$, interface probes identify the effective inputs and per-encounter
gains $\Delta_j$, and protocol checks screen the observation plan. The
completed path table supports per-path (Level 1) predictions, and an
aggregate (Level 2) prediction is issued only when all active paths are
enumerated, their $q_j$ and $\Delta_j$ identified, the gains stable, and
interactions specified or negligible; otherwise the aggregate is declared
unidentified.
Predictions are logged with curve class, applicable range and failure
assumption before execution. After execution, a nonzero composition
residual diagnoses an interaction or missing path, and any explanation
outside the frozen checklist is recorded as an audit failure. The path
table shows example entries with their checklist codes.}
\label{fig:framework}
\end{figure}

The decomposition turns an aggregate scaling curve into a set of
testable parts (Fig.~\ref{fig:framework}). Structural quantities such as
re-encounter and message coverage are derived from code. The responses of
language-model agents cannot be read from their weights and are therefore
measured with controlled probes. An aggregate prediction is made only
when the active pathways have been enumerated, their gains are known and
stable over the target range, and important interactions have been
specified or ruled out. A residual between the path sum and the observed
aggregate signals an interaction or a missing pathway.

A forecast must also exclude changes created by the experiment rather
than by the mechanism. The audit therefore checks that an outcome has the
same meaning at every population size, that the mechanism can activate,
and that the observation window is long enough for the relevant dynamics.
It also distinguishes information about a named individual, whose
coverage commonly shrinks with population, from type-level or public
information that may remain visible. These checks produce two outputs:
path-level predictions whenever structural exposure is known, and an
aggregate prediction only when the behavioural responses are known as
well. The complete checklist and pseudocode are given in Supplementary
Information.

%% file: sections_nc/results_single_terms.tex
\subsection{Single structural terms decide survival}

\begin{figure}[tbp]\centering
\includegraphics[width=0.95\linewidth]{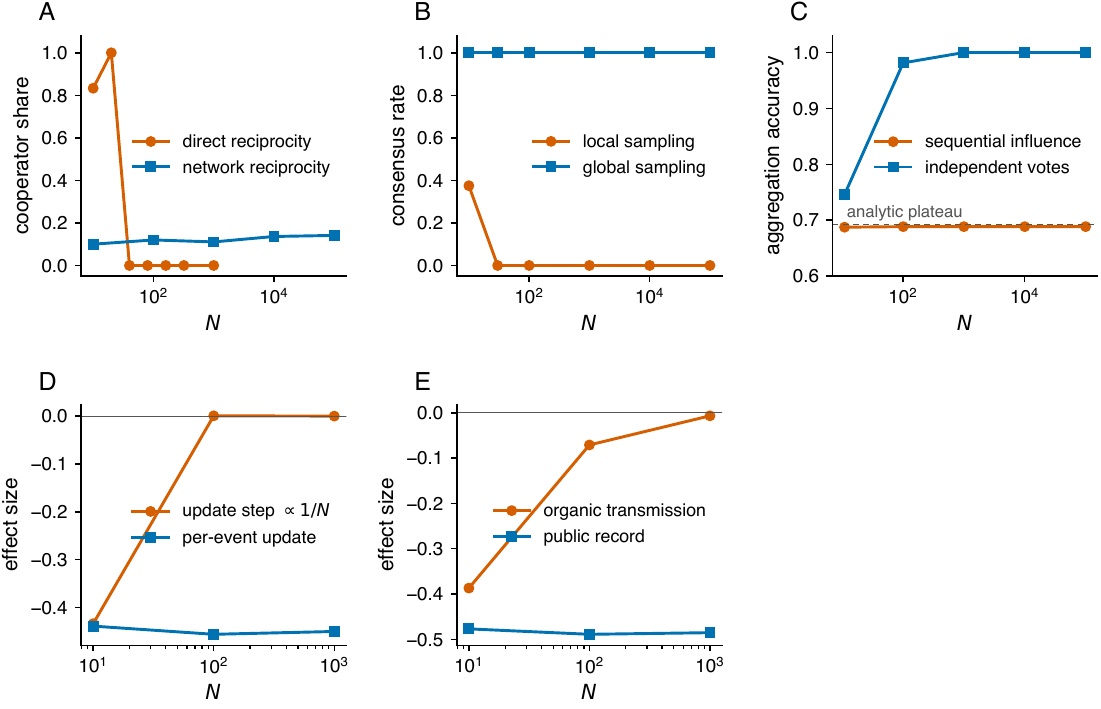}
\caption{Single structural terms set the scale class in rule-based
societies. (A) Direct reciprocity collapses where measured re-encounter
falls below threshold, while network reciprocity on fixed degree stays
flat. (B) Consensus under global sampling holds at every size, while the
same update rule under local sampling on a ring freezes from roughly
$N=30$ upward. (C) Independent aggregation improves with population,
while sequential social influence pins accuracy near the analytic plateau
of $0.692$. (D) A reputation mechanism whose update step carries a $1/N$
dilution factor loses its effect as the population grows, while the
per-event variant holds. (E) Replacing organic transmission with a public
record flips the scale class of the same gossip mechanism with individual
responses held fixed. Points are means over seeds (Methods).}
\label{fig:rule}
\end{figure}

We first tested the audit in rule-based societies, where every decision
rule is visible and a prediction can be checked directly against the code.
Four paired experiments kept the behavioural mechanism fixed while
changing one structural feature (Methods, Fig.~\ref{fig:rule}). Direct
reciprocity collapsed between $N=20$ and $N=40$, where random partners
stopped meeting often enough to sustain cooperation. Network reciprocity
on a fixed-degree lattice instead held a stable minority of cooperators,
unchanged from $N=10$ to $10^{5}$, because clusters of neighbours
continue to meet regardless of population. The pair differs in scale
class, collapse against flatness, not in who cooperates at small sizes. Likewise, global sampling preserved consensus while
local sampling on a ring froze into separate domains, and independent
aggregation improved with population while sequential social influence
limited the number of independent signals to about four
\cite{lorenz2011social}. In each case, changing the predicted bottleneck
restored the mechanism: fixing partners restored cooperation, adding
long-range links restored consensus, and adding independent information
restored aggregation accuracy.

A more direct intervention changed a single term in one update equation.
When a reputation update was divided by population size, its effect fell
from $-0.43$ to zero as the population grew. Removing that factor while
leaving the rest of the mechanism unchanged held the effect near $-0.45$
(Fig.~\ref{fig:rule}D). Population size therefore did not weaken the
agents themselves. It weakened the quantity delivered to them.

The same distinction separates an institution from the people who use it.
In an organic gossip system, a victim informed four randomly selected
agents. As population increased by two orders of magnitude, the share of
decisions carrying relevant gossip fell from $0.75$ to $0.013$, and the
sanction effect almost disappeared. Replacing private transmission with a
public record kept visibility near $0.93$ and the effect near $-0.48$
(Fig.~\ref{fig:rule}E). Individual responses were unchanged. The public
record altered only the scale of information coverage.

These experiments also expose two common misreadings of scale. A
mechanism measured at a ceiling can appear ineffective at every size, and
a fixed time horizon can make a slow system look as though it has failed.
After the saturated punishment experiment was moved into its dynamic
range, its effect was flat across populations. Finally, the direction of
an effect and its scaling are different properties. Reversing how victims
responded to punishment reversed the sign of the effect but left the same
coverage decay underneath. The audit predicts this decay. The
implementation determines whether the delivered effect is beneficial or
harmful.

%% file: sections_nc/results_product_law.tex
\subsection{Message reach and lifetime set the failure scale}

The audit can predict not only whether a mechanism decays, but where the
decay begins. In the gossip system, a report reaches $f$ people and remains
available for an effective lifetime $\taueff$. Their product is the
mechanism's information budget. The predicted crossover therefore obeys
$\Nc \propto f\,\taueff$: once the population substantially exceeds this
budget, too few decisions carry relevant gossip.

\begin{figure}[tbp]\centering
\includegraphics[width=0.8\linewidth]{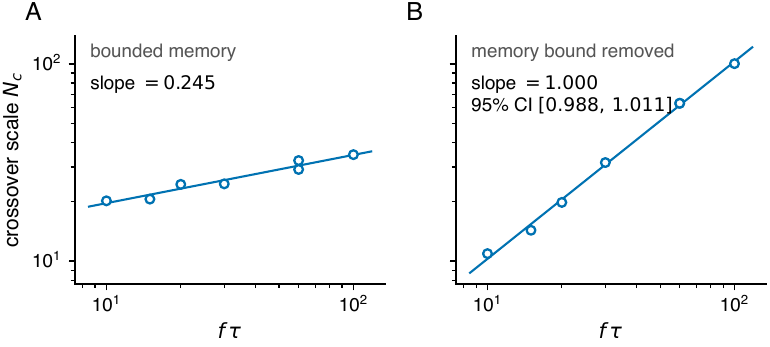}
\caption{Failure scale against the product of fan-out and mark lifetime
$f\tau$. (A) In the first design, which used memory bounded at twenty
marks and a relative crossing threshold, the exponent fitted against the
raw lifetime is $0.245$, outside the registered band. The bound can
truncate the effective lifetime $\taueff$ below $\tau$. (B) In the amended
memory-unbounded design, which uses fixed absolute crossing thresholds and
has $\taueff=\tau$, the exponent is $1.00$ with bootstrap intervals inside
$[0.97, 1.04]$ over ten seeds at three registered thresholds.}
\label{fig:exponent}
\end{figure}

The experiments recovered this relation quantitatively
(Fig.~\ref{fig:exponent}). With unbounded memory, the fitted exponent of
the crossover against $f\tau$ was $1.00$, with bootstrap intervals inside
$[0.97,1.04]$. Two systems with different fan-out and lifetime but the
same product failed at the same measured population of $63.1$. A finite
memory design using a relative crossing threshold yielded a compressed
exponent of $0.245$ and failed its registered test. Report turnover
provides a plausible account: new reports displace old ones and truncate
the effective lifetime, approximated by
$\taueff=(1/\tau+\lambda/M)^{-1}$. A separately registered amendment
removed the memory bound and used a fixed informed-fraction threshold,
and it yielded the exponent $1.00$. Because both features changed, this comparison
does not by itself show that correcting the bounded data by $\taueff$
recovers the law. It shows that the product law holds in the amended design
where retained and nominal lifetimes coincide.

%% file: sections_nc/results_formats.tex
\subsection{Language-model agents respond to information formats}

\begin{figure}[tbp]\centering
\includegraphics[width=0.6\linewidth]{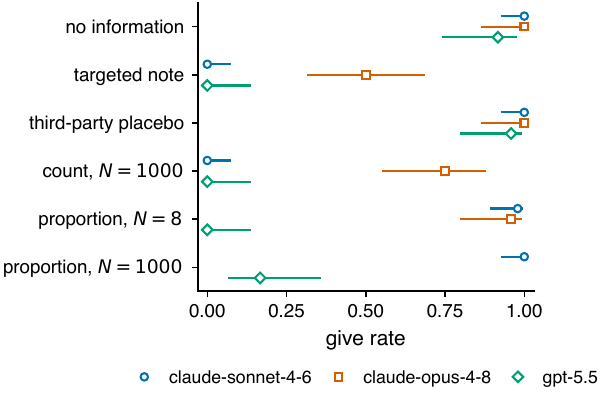}
\caption{Single-decision probe rates by cell and engine. Points are give
rates with $n$ between 24 and 48 decisions per cell, and horizontal bars
are 95\% Wilson intervals. A targeted report about the current partner
produces sanctioning, an identically worded report about a third party does
not, and count-format reports act without detected dependence on the
stated population size. The panel shows the six primary probe cells on
the three engines with per-cell reruns. The remaining engines and cells
are tabulated in Supplementary Information. Engine labels and run dates
are listed in the Supplementary Information model manifest.}
\label{fig:probes}
\end{figure}

For a language-model society, code reveals what information is shown to an
agent but not what the agent uses. We therefore isolated one decision at a
time: each probe asks an agent to decide whether to give part of its
endowment to a partner in a donation game (Methods). In the primary
engine, an agent with no information gave to its
partner, whereas a single report that the current partner had previously
kept a donation produced complete sanctioning. The same report about a
third party produced no detected effect (Fig.~\ref{fig:probes}). The response was
therefore tied to the identity of the current partner rather than to a
general negative tone. Its direction survived five independent
rewordings and appeared on five of six tested engines, although its
magnitude differed. These probes measure the behavioural gain of a
pathway without requiring the full social process to ignite.

\begin{figure}[tbp]\centering
\includegraphics[width=0.85\linewidth]{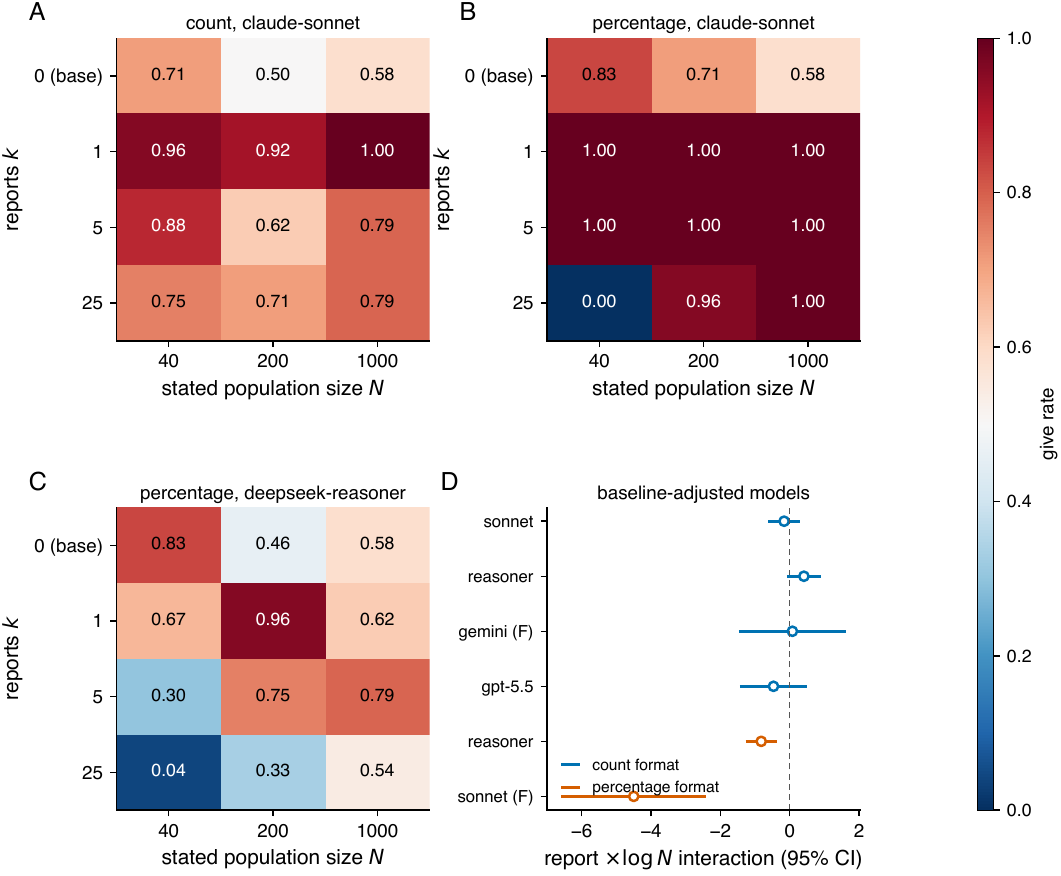}
\caption{Count and percentage formats lead to different stated-size dependence. (A) Dose lattice in the
count format on the primary engine, give rates with $n=24$ decisions per
cell. Response tracks the number of reports, while no stated-size
dependence was detected in this count-format lattice. Anti-diagonal cells
state identical proportions. (B, C) The same lattice with explicit
percentages on two model families. The stated size now acts, through a
sharp threshold on one engine and a graded curve on the other.
(D) Baseline-adjusted report $\times \log N$ interactions with 95\%
confidence intervals across all six lattices, a post hoc analysis
specified after the registered per-lattice fits (Methods). Count-format
intervals contain zero on every engine, while both percentage-format
interactions are negative and exclude zero. F marks Firth fits under
separation.}
\label{fig:dose}
\end{figure}

The representation of the evidence determined which quantity entered the
decision. We crossed the number of negative reports with a stated
population of $40$, $200$ or $1000$ (Fig.~\ref{fig:dose}). When the prompt
gave a count, for example ``five participants reported this
partner'', responses changed with the number of reports, but we detected
no additional effect of the stated population size. This result was
reproduced on further engines under criteria fixed in advance, with one
engine responding only to the presence rather than the number of reports.
The lattice wordings are deliberately mild so that responses stay away
from ceilings, which is why report cells can sit above the uninformed
baseline. The registered quantities are the dose and size slopes within
each lattice, where the wording is constant, rather than the
report-versus-baseline contrast.
By contrast, when the report made the percentage explicit, both
model families tested used population size: one showed a sharp threshold
and the other a graded response. The percentage sentence still contains
the count, so the added dependence tracks the added percentage. Cells describing the same percentage
then produced similar behaviour even when their counts and population
sizes differed.

The contrast is visible in baseline-adjusted models across all six
lattices. The report-by-population interaction was indistinguishable from
zero for every count-format lattice and negative for both
percentage-format lattices (Fig.~\ref{fig:dose}D, full coefficients in
Supplementary Information). Because the uninformed baseline itself can
drift with stated size, the count result is an absence of detected
size-dependence rather than proof of equivalence. Within that scope, the
finding is consequential: two mathematically equivalent descriptions of
social evidence can make the same mechanism appear scale-stable or
scale-sensitive.

We next asked whether gains measured in isolated decisions survive inside
a society. A full mechanism combined language-model decisions with
mechanical bookkeeping of message transmission and one scripted defector
to initiate the process. Away from the known final round, the measured
path contributions summed to the aggregate effect with zero residual at
both tested population sizes. The social process nevertheless had very
different reach. At $N=8$, it triggered ten sanctions and generated
second-order cascades. At $N=64$, it did not reach a single decision in
$945$ opportunities. Scale failure here was not a subtle change in
attitude: the mechanism stopped touching behaviour at all.

%% file: sections_nc/results_directions.tex
\subsection{Directions persist while absolute levels drift}

Commercial language models can change even when users call the same
product name. We therefore separated directional findings, whether one
condition raises or lowers a response, from absolute levels such as a
baseline rate or threshold location. Across repeated tests, the
targeted-report, third-party-placebo and count-format conclusions persisted
across model snapshots. A report about the current partner continued to
reduce giving, a third-party report remained a placebo, and the
count-format response retained its lack of detected size-dependence.

This stability was not universal. Absolute levels were less stable. The
uninformed giving rate moved from roughly $0.6$ to $0.93$ within days
under an unchanged model string, and a strong response to
percentage-framed evidence on one snapshot was absent on another. A
confirmatory rerun preserved the targeted and placebo contrasts but failed
its registered baseline threshold. We therefore attach model identity and
run date to every reported level. For language-model societies, the more
reproducible object can be the direction of a controlled contrast rather
than a timeless behavioural constant. That stability must nevertheless be
established for each interface.

%% file: sections_nc/results_prospective.tex
\subsection{Predictions generalize and failures narrow the claim}

We tested whether the audit could travel beyond the systems from which it
was developed. Adversarial language-model agents (Methods) first proposed
six counterexamples,
including spreading cascades and selection effects that could compensate
for declining local coverage. Each produced a measurable diagnostic
within the audit, while the exercise added explicit checks for
scale-comparable outcomes and insufficient time horizons.

\begin{figure}[tbp]\centering
\includegraphics[width=0.95\linewidth]{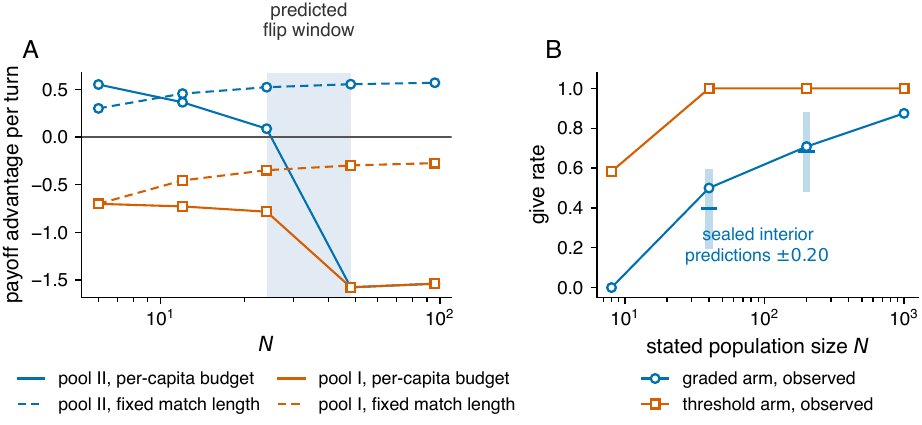}
\caption{Predictions filed in advance and their outcomes. (A) Third-party
cooperation library on a frozen parameter grid. Lines are the filed
predictions and open markers the observed values, which agree exactly on
this deterministic system. The shaded band is the filed sign-flip window.
(B) Cross-model-family test filed before execution, with the report count
fixed at three. Curves are observed give rates for the graded and
threshold arms, and the blue ticks with bands are the two interior
predictions filed in advance with their $\pm 0.20$ bands, both met by the
graded arm. The threshold arm failed its registered flatness criteria.}
\label{fig:prospective}
\end{figure}

The audit was then applied to unmodified external systems. On a community
cooperation library \cite{knight2016open}, it predicted a sign change
between $N=20$ and $N=40$. A new parameter grid was frozen before
execution, and all twenty predicted values agreed with the deterministic
outputs, a check that the audit's derivation is correct rather than a
statistical test (Fig.~\ref{fig:prospective}A). Result-masked readings of published
models produced two valid cases and one voided case
\cite{haghrah2026scalability, de2026collective}. The audit recovered the
curve class and memory-to-population collapse of a bounded-memory model,
although it placed the absolute crossover too early by a factor of
two to two and a half. For a lattice of language-model agents, it correctly refused
an aggregate forecast until coupling and bias were separately measured.
These outcomes show both the reach of the structural classes and the
lower precision of absolute crossover locations.

A final prediction was filed before testing a second model family
(Fig.~\ref{fig:prospective}B). The engine with a graded response
followed the predicted monotonic curve and met both interior prediction
bands. The companion prediction that a threshold-style engine
would remain flat failed every registered criterion. Follow-up probes
showed that this engine read the explicit percentage through a
threshold, narrowing the earlier count-format finding rather than
changing it after the fact. All failed and inconclusive outcomes are
reported in Supplementary Information. In this framework, a failure is
not discarded: it marks the boundary of the mechanism that can be
predicted.

%% file: sections_nc/discussion.tex
\section{Discussion}

Artificial societies are attractive because they can be enlarged, copied
and intervened on in ways that human societies cannot. The same
flexibility creates a basic interpretive problem: adding agents changes
who meets whom, what information remains visible and how an outcome is
measured. Our results show that these changes can be separated before a
large simulation is run. Re-encounter, information coverage and available
time determine whether a pathway reaches decisions. Controlled probes
determine what agents do with the information once it arrives. Their
combination predicts the fate of the mechanism rather than merely
describing the curve that appears afterwards.

The most general discovery is that scale dependence need not originate in
the decision-maker. The same agents and behavioural rule produced
opposite scale outcomes when one structural term changed. A public record
preserved gossip by maintaining visibility, whereas private transmission
lost coverage. More broadly, institutions can be understood as devices
that preserve information and coordination as populations grow
\cite{milgrom1990role, kandori1992social}. This interpretation is
consistent with human experiments in which reputation remains effective
when scores are visible to all participants \cite{wu2020cooperation}.
Scale invariance is therefore not necessarily an intrinsic property of a
social mechanism. It can instead be an achievement of the surrounding
information system.

Language-model agents add another layer because natural language is part
of the mechanism. Counts and percentages that encode the same underlying
ratio led to different population-size responses. This result connects a
design choice in the prompt to a population-level prediction. Human
decision-makers also show ratio bias, denominator neglect and nonlinear
responses to the number of sources \cite{kirkpatrick1992cognitive,
denes1994conflict, reyna2008numeracy, latane1981psychology}, but such
parallels do not make a language model a human participant. Instead, they
show why the cognitive interface must be measured for each agent
population. The observed drift across model snapshots strengthens this
point: controlled directions were more stable than baseline levels, so
behavioural rates from an agent society should be reported with model
identity, date and information format.

The same accounting can organize observations outside artificial
societies without turning them into validation data. As an illustrative
case, we applied the checklist to peer review at ICLR (the International
Conference on Learning Representations), one of the largest
machine-learning conferences, whose submissions grew roughly forty-fold
in ten years as the field expanded. Its public records show deadline
concentration, an expanding stock of coordination devices, and an apparent
collapse of revision activity created by a change in the public record.
The deadline concentration is largely absent at TMLR (Transactions on
Machine Learning Research), a journal in the same field whose reviews run
on per-paper rolling deadlines rather than one synchronized deadline,
which makes it a natural contrast for deadline-driven behaviour.
These descriptive results, reported in Supplementary Information, show
that structural coverage, compensation and measurement artifacts are
recognizable in a real institution. They do not establish causal effects
or test whether language-model agents reproduce human groups.

The framework has clear limits. Absolute crossover locations should be
treated as coarse forecasts: in the external bounded-memory case, the
predicted crossover was a factor of two to two and a half lower than the
observed one. The framework can therefore guide the choice of population
sizes to test, but precise threshold estimates will require a more
detailed model of memory turnover. The language-model findings concern one
donation-game family, the tested wordings and the named model
snapshots. Strong interactions between pathways prevent an aggregate
prediction, and an unknown gain produces an unidentified verdict rather
than a forecast. A nonzero composition residual is not merely a failure
signal: it localises the interactions that a linear path sum cannot
carry, and marks where studies of emergent coordination should look
next. Predictions were logged internally before execution but
were not independently preregistered. The framework is falsifiable on its
own terms: a filed prediction that fails without an explanation inside the
frozen checklist is an audit failure, a case requiring a new category
rejects the checklist's completeness, and unidentified verdicts are
recorded rather than counted as successes. The field application concerns
one institution observed without intervention. These restrictions define
where the present evidence ends and where direct tests in human
populations remain necessary.

The audit changes the question asked of a large agent society. Instead of
waiting to see whether an aggregate pattern changes with population, it
asks which route carries the mechanism, how that route scales, and
whether the agents use the information it delivers. Those quantities are
measurable before the full run. This makes population scale a property to
be predicted and designed for, rather than a robustness check performed
after an expensive simulation has already produced a result.

%% file: sections_nc/methods.tex
\section{Methods}
\label{sec:methods}

\subsection{Audit procedure and checklist}
The audit uses a frozen, closed checklist fixed before the external test
batch and proceeds in four steps. First, it enumerates the routes through
which a mechanism can affect a decision. Second, it derives how often each
route reaches a decision as population size changes. Third, it measures
the behavioural response delivered by that route. Fourth, it combines
routes only when their gains and interactions are identified; otherwise
it retains the path-level predictions and marks the aggregate as
unidentified. Before a prediction is filed, the checklist also screens
for changes in outcome meaning, mechanism activation, observation time,
capacity, saturation and compensating mechanisms.

Predictions cite checklist codes and state a curve class and an applicable
range. Explanations outside the frozen checklist count as audit failures.
Compensation claims require quantitative diagnostics, with parallel paths
identified and their ablations registered in advance. The adversarial
attacks and the result-masked audits were performed by
clean-session language-model agent auditors isolated from the result
data, with access limited to methods-only briefs and, for the red team,
to the frozen checklist. They are not an independent research team, the
authors adjudicated every outcome against the checklist, and auditor
recognition of the research lineage voided one case.
Retrodictions of published results are
performed under result masking, with contamination self-reports treated as
a weak filter, since training-data exposure of published results cannot be
excluded. The full checklist, the probe materials with their registered criteria
and adjudications, and the twelve-entry failure ledger are in
Supplementary Information.

\subsection{Rule-based experiments}
All rule-based simulations use seeded NumPy code with explicit seeds and at
least three seeds per configuration, with populations from $N=10$ to
$10^{5}$ as stated per experiment. The paired designs hold the update rule
fixed and vary one term, the sampling source for consensus, the presence of
$1/N$ in an update step for reputation dilution, the information routing
for the gossip pair, and the coverage generator for the institutional
comparison (five seeds, fifty rounds, $N \in \{10, 100, 1000\}$). In the causal
interventions the restored arms sit at the ceiling, which is harmless
since the verdicts require only recovery while the degraded baselines
remain interior. The
failure scale law uses ten seeds per configuration, three registered
coverage thresholds ($0.2$, $0.3$, $0.4$), and least squares fits of
$\log \Nc$ on $\log(f\,\tau)$ with bootstrap intervals over seeds, run on
the amended memory-unbounded design in which $\taueff=\tau$, using a fixed
absolute informed-fraction threshold. The bounded first design used a
relative half-plateau threshold. Its registered failure at exponent
$0.245$ and the truncation formula
$\taueff=(1/\tau+\lambda/M)^{-1}$ are reported in Results. The transfer and
prospective tests run
on the unmodified Axelrod-Python library \cite{knight2016open} with frozen
parameter grids and predictions filed before execution.

\subsection{Language-model probes}
Probes present a donation game vignette and elicit a single give or keep
decision, with $n=24$ decisions per cell unless stated, frozen wordings,
and resume-safe logging. Later runs record the full provider envelope
including the resolved model identity per call. The earliest probes
predate envelope logging and carry the most specific engine label
available, as detailed in the Supplementary model manifest.

Engines span five model families (Claude
Sonnet 4.6 and Sonnet 5, Claude Opus 4.8, GPT-5.5, Gemini 3.5 Flash, and
DeepSeek v4-flash accessed with its reasoning mode on and off over fixed
weights). Three of the six engines were accessed through vendor agent
CLIs and three through direct APIs, and the directional and format
results replicate across both access routes. Every reported comparison is
made within a single model snapshot, and replication across snapshots is
itself reported in Results.

The count lattice crosses $k \in \{1,5,25\}$ with stated
$N \in \{40,200,1000\}$ plus $k=0$ baselines, with per-engine wording
calibration to place the anchor cell in the interior. The percentage
lattice states the arithmetic-consistent percentage inside a directly
worded report. The calibrated count wordings are milder than the
percentage wording, so the lattices are not matched sentence pairs, and
the format contrast rests on within-lattice slopes, in which the wording
is constant, together with the matched core probe cells whose count and
percentage sentences differ only by the inserted percentage. Calibration used a separate pilot batch of twelve
decisions per candidate wording, pilot calls enter no reported analysis,
and every lattice cell, including the calibration point, was collected
fresh in the confirmatory runs. Per-engine calibration supports interface
measurement within each engine rather than same-stimulus comparison across
engines.

The cross-model-family test fixed the report count
at three, varied stated population over $\{8,40,200,1000\}$, and filed
monotonicity, two anchor values and two interior values from a logistic
interpolation with bands of $\pm0.20$ before any data existed. The composed
system uses one scripted defector and mechanical note transmission, with
registered verdicts adjudicated inside the operating window and final-round
cells reported as failures. Temperature is provider-fixed on the primary
engine, and a sweep at $t \in \{0,0.7,1.0\}$ on another engine reproduced
all cells identically.

\subsection{Field study}
Data are public OpenReview records retrieved with the official client under
authenticated access, at most two requests per second, with resume-safe
provenance logs. Submissions are enumerated by invitation as unique root
forums with a seven-category status breakdown, reviews are official reviews
on those forums, and revisions are public edit records classified by
signature, with reviewer revisions defined as edits signed by the review's
own reviewer pseudonym more than $120$ seconds after review creation.
Trusted timestamps (tcdate, tmdate) are used throughout, since
user-editable dates are not relied on. Enumerated counts reconcile against
official totals, including an exact match of $11{,}672$ root forums for
2025, with remaining gaps documented in Supplementary Information. The 2025
area split used a fixed seed (20260719), development-half analyses preceded
a single validation run with three criteria set in advance, and the 2026
incident windows were fixed from the official timeline. Device coding used
only public policy documents across 2017 through 2026, with fifty devices
in five categories, per-device sources, and a persistence rule stated in
Supplementary Information. Coding was performed by one coder, and the
released per-device source list supports independent recoding.

\subsection{Ethics and data use}
This study recruited no participants and performed no interventions. The
field component is a secondary analysis of publicly accessible,
human-generated peer review records retrieved from the platform's public
API in accordance with its terms of use
(\url{https://openreview.net/legal/terms}). The authors had no interaction
with reviewers and accessed no private information. Review comments on the
platform are published under a CC BY 4.0 licence and metadata under CC0.
No review text is quoted or redistributed, anonymous identities are not
linked across papers or to natural persons, and only aggregate counts and
timing distributions are reported. The randomized feedback intervention described
in Supplementary Information was run by the conference itself under its own institutional
approval, and this study neither participated in it nor analyses it
causally.

\subsection{Statistics and reproducibility}
The statistical unit of every probe analysis is a single API call in an
isolated session with no shared context. Calls were treated as
conditionally independent given the engine, wording and sampling
configuration, and the
randomness is the provider's sampling randomness rather than sampling
from a population of models. All inference is therefore scoped to the
tested snapshot and prompt distribution, which is the same scope as the
directional-versus-level distinction reported in Results, and the
five-rewording robustness runs bound sensitivity to any single wording.
Per-cell samples of $n=24$ are powered for the large contrasts the probes
target, rates near zero against rates near one, and lattice-level
inference pools $216$ decisions per engine in the registered models
rather than testing single cells.

All probe rates are proportions of valid parsed decisions with exact $n$
reported per cell. Infrastructure failures were retried to completion and
the superseded records retained in the raw logs. Dose lattice analyses are
binomial logistic regressions (generalized linear models, GLMs) of the
keep decision on $\log k$ and $\log N$. These regressions are the test
of which quantity enters the decision: a nonzero coefficient on $\log k$
means the reported count is used, and a nonzero coefficient on $\log N$
means the stated population size is used ($n=216$ decisions per engine
per lattice, $215$ on one engine),
two-sided Wald tests, with Firth's penalised likelihood under separation,
and the ratio-reading contrast tests
$H_0\!:\beta_{\log k} = -\beta_{\log N}$ (two-sided). For the
primary count lattice the exact values are $\beta_{\log k}=+0.48$
($z=3.17$, $p=0.0015$) and $\beta_{\log N}=0.00$ ($z=0.00$, $p=1.00$).
For the percentage lattice they are
$+0.66$ ($z=5.29$, $p=1.2\times10^{-7}$) and $-0.48$ ($z=-3.95$,
$p=7.8\times10^{-5}$) on one engine with contrast $z=1.15$
($p=0.25$), and Firth $+4.8$ and $-4.1$ on the other with contrast
$z=0.68$ ($p=0.50$). For the count lattice on further engines the
$\beta_{\log k}$ dose tests give $z=7.49$ ($p<10^{-10}$), $z=3.05$
($p=0.0023$), and $z=1.10$ ($p=0.27$, the gated engine), with
$\beta_{\log N}$ intervals $[-0.17,+0.33]$, $[-0.36,+0.10]$ and
$[-0.19,+0.26]$.

The baseline-adjusted models reported in Results add the
$k=0$ no-information cells and fit
$\mathrm{keep} \sim 1 + R + R\log k + \log N + R\log N$ per lattice and
engine, where $R$ indicates report presence and $R\log k$ carries the dose
among report cells. The reported quantity is the $R\times\log N$
interaction with two-sided Wald 95\% intervals, Firth's penalised
likelihood where the plain fit separates, and no multiplicity correction
across the six lattices since the six intervals are reported in full. This
analysis was specified after the registered per-lattice fits and is
reported as post hoc. Full coefficient
tables are in Supplementary Information.

The failure scale law reports least squares fits with
bootstrap intervals from $1{,}000$ resamples over ten seeds at three
thresholds. Field clustering shares are proportions with exact
denominators ($29{,}045$ development and $17{,}703$ validation reviews),
and area correlations are two-sided Spearman rank correlations with tie
handling, $\rho=-0.49$ ($p=0.15$, $n=10$ areas) and $\rho=+0.18$
($p=0.60$, $n=11$ areas). Language-model calls total roughly $6{,}600$. Every level-type quantity
carries the most specific available engine label and a run date in the
project ledger, and fully resolved provider identities are available for
the later envelope-logged runs.

%% file: sections_nc/availability.tex
\section*{Data availability}
All raw experimental outputs, including JSON and JSONL files containing one record per language-model decision with its engine label and, where logged, the resolved provider envelope, together with the field-study reconciliation counts and provenance logs, device coding matrix, and prediction ledger, are available at \url{https://github.com/wuzengqing001225/scale_limits_agent_societies} and archived at Zenodo (\url{https://doi.org/10.5281/zenodo.21532739}).
The field study also analysed publicly accessible OpenReview records. The raw OpenReview records are not redistributed in the repository; they can be retrieved from \url{https://openreview.net} through the OpenReview API, subject to the platform's Terms of Use, using the extraction code provided in the repository.

\section*{Code availability}
All simulation, probe, extraction and analysis code, including the scripts
that regenerate every figure and every Supplementary table from the
released data, are available in the same repository
(\url{https://github.com/wuzengqing001225/scale_limits_agent_societies})
and archived at Zenodo (\url{https://doi.org/10.5281/zenodo.21532739}).

%% file: sections_nc/backmatter.tex
\section*{Acknowledgements}
C.X. discloses support for the research of this work from JSPS KAKENHI
(JP23K17456, JP23K28096, JP25H01117 and JP26K03246) and JST CREST
(JPMJCR22M2). Z.W. discloses support for the research of this work from
JST BOOST (JPMJBS2402).

\section*{Author contributions}
Z.W. conceived the study, performed the experiments and analyses, and wrote
the manuscript. C.X. supervised the research and revised the manuscript.

\section*{Competing interests}
The authors declare no competing interests.

%% file: si_body.tex
\section{The audit checklist}

The checklist is closed. Predictions cite the codes below, are filed before
results together with a curve class and an applicable range, and an
explanation from outside the list counts as an audit failure. Revisions
create a new version, and predictions are adjudicated under the version
they were filed under. The checklist used for every result in the paper was
fixed before the external test batch. Tables~\ref{tab:si-checklist-a}
and \ref{tab:si-checklist-b} give the closed code list with each code's
rule, diagnostic and permitted use. The paragraphs that follow state the
standards governing their application.

\begin{table}[h]
\centering\footnotesize
\caption{The closed checklist, part one: protocol checks (C1--C5) and
input identity rules (E1--E4). Every prediction cites codes from this
list, and an explanation outside it is an audit failure.}
\label{tab:si-checklist-a}
\setlength{\tabcolsep}{3pt}
\begin{tabular}{@{}p{0.8cm}p{2.6cm}p{5.8cm}p{3.4cm}@{}}
\toprule
Code & Name & Rule and diagnostic & Use in predictions \\
\midrule
C1 & Scale-comparable observable & The observable must have an
$N$-comparable interpretation. Extensive counts and whole-population
conjunctions, such as $P(\text{all agree}) = p^{N}$, carry mechanical
size dependence and require normalisation or a separate null model &
prediction undefined if violated \\
C2 & Activatability & Mechanism must activate at the reference scale or
under exogenous seeding. Zero ignition only at large $N$ is a positive
A2-class result, zero ignition everywhere is an invalid test & validity
gate \\
C3 & Causal neighbourhood & Neighbourhood within the measurement horizon
must clear the boundary and stay locally isomorphic across $N$. On rings
this reduces to $N \gtrsim 2rT$. Globally coupled systems need separate
argument & validity gate for small $N$ \\
C4 & Endgame isolation & Final rounds of known horizons excluded or
reported separately & measurement window \\
C5 & Dynamic range & No ceilings or floors. Saturated cells are retuned
and excluded from adjudication & measurement window \\
\midrule
E1 & Identity-indexed input & Decision function indexed by the identity of
the object & routes to A2 coverage audit \\
E2 & Type or density input & Type frequency does not dilute mechanically
with the denominator, but its generating dynamics remain subject to
A4/A5 & waives A2 only \\
E3 & Count versus proportion & The agent's actual weighting of numerator,
denominator and format is measured by probes & fixes the effective input
and supplies $\Delta_j$ \\
E4 & Public record versus private transmission & A public record is only a
candidate for B3. Measured visibility must decouple from $N$ &
precondition for B3 \\
\bottomrule
\end{tabular}
\end{table}

\begin{table}[h]
\centering\footnotesize
\caption{The closed checklist, part two: degradation classes (A1--A5),
correction terms (I1--I3), and compensation channels (B1--B4). Correction
terms shift the absolute position of a turning scale and never justify a
decay prediction on their own. Compensation channels require the stated
quantitative diagnostic at the time of invocation.}
\label{tab:si-checklist-b}
\setlength{\tabcolsep}{3pt}
\begin{tabular}{@{}p{0.8cm}p{2.6cm}p{5.8cm}p{3.4cm}@{}}
\toprule
Code & Name & Rule and diagnostic & Use in predictions \\
\midrule
A1 & Re-encounter probability & Pair re-interaction probability, under
random matching $\sim 1/(N{-}1)$. Diagnostic: count the pair generator &
decay or transition of $q_j$ \\
A2 & Targeted coverage & Probability that a decision point holds
information about the current object, $\sim \lambda\,\taueff/N$.
Diagnostic: measure the informed fraction & decay or transition of $q_j$ \\
A3 & Monitoring reach & Probability that behaviour is observed by a
potential sanctioner. Diagnostic: count observation edges & decay or
transition of $q_j$ \\
A4 & Time budget & Convergence time $\tau(N)$ against a fixed horizon $T$.
Diagnostic: measure $\tau(N)$, discriminate by extending $T$ & decay or
transition of $q_j$ \\
A5 & Update-step dilution & A $1/N$ factor inside the update step of a
global aggregate. Diagnostic: read the update rule & decay or transition
of $q_j$ \\
\midrule
I1 & Memory-flux lifetime & $\taueff = (1/\tau + \lambda/M)^{-1}$,
harmonic mean-field approximation under independent hazards & shifts
turning scale only \\
I2 & Capacity and throughput & Finite processing or storage limits &
shifts turning scale only \\
I3 & Behavioural saturation & Saturation of the behavioural mapping,
distinct from measurement saturation (C5) & shifts turning scale only \\
\midrule
B1 & Supercritical spreading & Compute or measure $R_0$. Coverage capped
rather than $\sim 1/N$ & flat, with diagnostic \\
B2 & Selection amplification & Stationary share converges to an
$N$-independent value & flat, with diagnostic \\
B3 & Institutional coverage restoration & Measured visibility or query
rate decoupled from $N$ (E4 precondition) & flat, with diagnostic \\
B4 & Parallel $N$-independent path & Path identified in the path table in
advance. Ablation registered in advance, execution may follow &
conditional, with registered ablation \\
\bottomrule
\end{tabular}
\end{table}

\paragraph{Two-level output.} Level 1 predicts structural quantities $q_j$
per causal path without using the target system's scale results, from code,
mechanism descriptions, and independent interface probes. Level 2 predicts
the aggregate $E(N)$ only when every active path's gain is derived or
independently measured and shown stable over the target range, or measured
at several sizes. Otherwise the verdict is aggregate unidentified. Each
prediction carries a curve class among flat, monotone decay, monotone
growth, non-monotone or transition, and conditional, together with an
applicable range, a turning region when computable, cited checklist codes,
and the assumption most likely to fail.

\paragraph{Protocol checks (C1--C5).} Observables must be
scale-comparable. Extensive counts and whole-population conjunctions
carry mechanical size dependence and require normalisation or a separate
null model. Mechanism
activatable at the reference scale or under exogenous seeding, where
non-ignition at large $N$ alone is a result and non-ignition everywhere is
an invalid test. Causal neighbourhoods must clear the boundary and remain
locally isomorphic across sizes, which on rings reduces to a lower bound on
$N$ of order twice the interaction radius times the horizon, while globally
coupled systems need separate argument. Final rounds of known horizons are
excluded or reported separately. Measurements must sit away from ceilings
and floors, and saturated cells are retuned and excluded.

\paragraph{Input identity rules (E1--E4).} Identity-indexed information
enters the coverage audit. Type and density information does not dilute
mechanically, though its generating dynamics still require audit. Count
versus proportion consumption is settled by probes. Public records are
candidates for coverage restoration only after measured visibility
decouples from $N$.

\paragraph{Path decomposition.} A mandatory table lists each path with
activation status, effective input, degradation class, gain source among
code, probe, and unknown, and the gain's sign and magnitude or the
response function from which the gain is derived. Any unknown gain halts
the audit at Level 1.

\paragraph{Degradation classes (A1--A5) and corrections (I1--I3).}
Re-encounter probability, targeted coverage, monitoring reach, time budget
against convergence time, and update-step dilution. Corrections cover the
memory-flux effective lifetime, capacity and throughput limits, and
behavioural saturation, and none of them justifies a decay prediction on
its own.

\paragraph{Compensation channels (B1--B4).} Supercritical spreading,
selection amplification, institutionalized coverage restoration, and
parallel $N$-independent paths. Every invocation requires a quantitative
diagnostic, and parallel paths must be identified in advance with ablations
registered in advance. Opposite-signed paths yield conditional verdicts
with explicit switching conditions.

\paragraph{Non-monotone standards.} A non-monotone or transition verdict
requires structure specified in advance, either opposite-signed paths with
a predicted crossing region or an explicit single-path non-monotone
equation with a turning region. Observed non-monotonicity without an
advance basis counts as an audit failure.

\paragraph{Retrodictions.} Audits of published results are performed under
result masking, with contamination self-reports treated as a weak filter.

\subsection{Audit procedure as pseudocode}

\begin{quote}\small
\textbf{Input:} mechanism implementation $M$, agent population $A$,
observation plan $O$, frozen checklist $C$.\\
\textbf{1.} Screen $O$ against the protocol checks of $C$; retune or
exclude any measurement at a ceiling, on a non-scale-comparable
observable, under a
known final round, inside a too-small causal neighbourhood, or under a
time budget shorter than convergence.\\
\textbf{2.} Enumerate the causal paths $j$ of $M$ from its implementation;
for each, identify the carrying quantity and classify its input under the
input identity rules of $C$.\\
\textbf{3.} For each path, derive $q_j(N)$ from the implementation, and
measure $\Delta_j$ with single-decision probes of $A$ in the formats that $M$
actually uses, with frozen wordings and interior calibration.\\
\textbf{4.} Complete the path table; if any gain is unknown, stop at
per-path predictions and return the verdict aggregate unidentified.\\
\textbf{5.} Otherwise predict $E(N)$ from
Eq.~(1) of the main text; check compensation channels of $C$ with
quantitative diagnostics before predicting decay.\\
\textbf{6.} Log every prediction with curve class, applicable range, cited
checklist codes, and the assumption most likely to fail, before results
exist; adjudicate outcomes against the logged text; report failures.\\
\textbf{Diagnostics:} a nonzero composition residual signals an
interaction or a missing path; out-of-checklist explanation of any outcome
is an audit failure.
\end{quote}

\section{Completed path tables}

Tables~\ref{tab:si-paths-rule} and \ref{tab:si-paths-llm} consolidate the
path tables of the main systems as completed during the audits. Structural
quantities, gains and curve classes were recorded before the corresponding
scale outcomes. The outcome column reports the adjudicated result and was
added after. The anchor pairs A to C formed the inducing set from which
the checklist was frozen and are marked accordingly. Their codes are the
retrospective classification under the frozen list.

\begin{table}[h]
\centering\scriptsize
\caption{Path tables, rule-based systems. Gains derive from code. Sources
in parentheses. Classes are flat, decay, or transition (decay crossing a
registered threshold).}
\label{tab:si-paths-rule}
\setlength{\tabcolsep}{2.5pt}
\begin{tabular}{@{}p{2.0cm}p{2.0cm}p{1.6cm}p{2.7cm}p{2.2cm}p{0.8cm}p{1.2cm}p{1.6cm}@{}}
\toprule
System & Path & Input (E) & $q_j(N)$ (source) & $\Delta_j$ (source) &
Codes & Filed class & Outcome \\
\midrule
Pair A, direct (anchor) & remembered re-encounter & partner identity (E1)
& $\sim 1/N$ under random matching (code) & full conditional defection
(code) & A1 & transition & collapse at $N{=}20$--$40$ at threshold \\
Pair A, network & neighbour clustering & neighbour payoffs &
$N$-independent local structure (code) & imitation of best neighbour
(code) & -- & flat & share stable near $0.1$ to $10^{5}$ \\
Pair B, local (anchor) & convergence within budget & neighbour opinions &
$\tau(N)$ outruns fixed $T$ (measured) & majority adoption (code) & A4 &
transition & frozen from $N \approx 30$ \\
Pair B, global & mean-field sampling & sampled opinions & $N$-independent
(code) & same (code) & -- & flat & flat to $10^{5}$ \\
Pair C, sequential (anchor) & informative decisions & public record &
saturates near four (measured) & cascade adoption (code) & -- & flat
below analytic anchor & pinned near $0.689$ vs $0.692$ \\
Pair C, independent & private signals & own signal & grows with $N$
(code) & majority count (code) & -- & growth & $0.75 \to 1.00$ \\
R4, diluted & reputation update step & aggregated reports & step
$\propto 1/N$ (code) & defection below threshold (code) & A5 & decay &
$-0.43 \to 0.00$ \\
R4, per-event & reputation update step & reports & $N$-independent (code)
& same (code) & -- & flat & holds near $-0.45$ \\
R6, organic & targeted coverage & identity marks (E1) & $\sim 1/N$
(measured $0.75 \to 0.013$) & multiplicative sanction (code) & A2 & decay
& effect $-0.387 \to -0.007$ \\
R6, board & institutional coverage & public record (E4) & visibility
$\equiv 1$ by construction & same (code) & B3 & flat & holds at $-0.48$ \\
R3$'$ gossip & targeted coverage & identity marks (E1) &
$\approx \min\{1, \lambda\taueff/N\}$ (code) & $p$-response to mark
(code) & A2, I1 & transition, $\Nc \propto f\taueff$ & exponent $1.00$
$[0.97, 1.04]$ \\
\bottomrule
\end{tabular}
\end{table}

\begin{table}[h]
\centering\scriptsize
\caption{Path tables, language model and external systems. Gains are
probe-measured on the stated engines. Give rates with $n$ per cell in the
result tables.}
\label{tab:si-paths-llm}
\setlength{\tabcolsep}{2.5pt}
\begin{tabular}{@{}p{2.1cm}p{2.0cm}p{1.8cm}p{2.5cm}p{2.5cm}p{0.9cm}p{1.2cm}p{1.5cm}@{}}
\toprule
System & Path & Input (E) & $q_j(N)$ (source) & $\Delta_j$ (source) &
Codes & Filed class & Outcome \\
\midrule
Count channel & targeted report & report count (E3, probed) & stated $N$
not consumed, no dilution path (probes) & give $1.00 \to 0.00$ on
targeted report. Dose $+0.48$ per $\log k$ (probes) & E3 & flat in
stated $N$ & no detected stated-size dependence, four engines \\
Percentage channel & stated ratio & explicit percentage (E3, probed) &
report always delivered in probes; its stated ratio $k/N$ falls with $N$
& threshold (Firth $+4.8/-4.1$) or graded ($+0.66/-0.48$) response to
the ratio (probes) & E3 & decay in stated $N$ & negative interactions,
both model families \\
Composed mechanism & ignition, transmission, sanction & identity marks
(E1) & fired at $N{=}8$. Zero ignition at $N{=}64$ over $945$ decisions
& probe values above & A2, C2, C4 & window-qualified composition &
residual zero in window. Whole-run verdicts failed on final round \\
Third-party library & within-match re-encounter & partner identity (E1) &
repeat turns per opponent fall under per-capita budget (code) & strategy
payoff structure (code) & A1, A4 & sign flip in $N{=}24$--$48$ & $20/20$
filed quantities exact \\
\bottomrule
\end{tabular}
\end{table}

\section{Rule-based model specifications}

Complete implementations with all constants are in the released code
(Code availability). This section specifies each model's state, update
order, and the single manipulated term of its pair.

\paragraph{Pair A, direct versus network reciprocity (\texttt{coop\_abm.py}).}
$N$ agents hold strategies TFT or ALLD (initialised uniformly at random),
benefit $3$ and cost $1$, and a bounded per-agent memory of $M=20$
(partner identity, last move toward the agent) written circularly. Each
generation plays $50$ rounds of random perfect matching followed by
payoff-based strategy updating, for $30$ generations. TFT cooperates
unless the remembered last move of the current partner was a defection.
The measured structural quantity is the remembered re-encounter rate.
The network arm is the classical spatial game: cooperators and defectors
on a ring with two neighbours, payoffs summed over neighbours, and each
agent imitating its best-performing neighbour, run for fifty rounds. The
registered intervention in the direct arm fixes partners (A1 restored).
The observable in both arms is the cooperator share.

\paragraph{Pair B, consensus versus segregation (\texttt{consensus\_abm.py}).}
$N$ agents hold opinions $\pm 1$ and adopt the majority of themselves plus
four sampled agents, for a fixed budget of $200$ sweeps. The manipulated
term is the sample source only, fixed ring neighbours at offsets
$\pm 1, \pm 2$ against fresh uniform global samples. Consensus is
unanimity. The discordant-edge fraction is recorded. The registered
intervention samples globally with probability $\varepsilon$ per update.
The surviving twin runs Schelling segregation on the same ring.

\paragraph{Pair C, information aggregation (\texttt{agg\_abm.py}).}
Binary ground truth, private signal accuracy $q=0.6$, $2{,}000$
replicate runs evaluated at checkpoints $N$ from $11$ to $10^{5}$
(odd, no ties). The independent arm counts a majority of private signals.
The sequential arm makes each agent adopt the sign of the running public
action difference whenever its magnitude is at least two, ignoring its
own signal (a cascade), and follow its signal otherwise. The analytic
plateau is $q^{2}/(q^{2}+(1-q)^{2}) = 0.692$. The measured structural
quantity is the number of informative, non-cascaded decisions. The
intervention gives each agent probability $\lambda$ of using its own
signal regardless of the record.

\paragraph{R4, update-step dilution (\texttt{r4\_dilution.py}).}
$N$ agents with fixed cooperation propensities drawn from $U(0.4, 0.9)$
and a reputation score in $[0,1]$ initialised at $1$. Partners are drawn
from ring offsets within distance two. Agents defect toward partners with
reputation below $0.5$ with probability $0.8$. After each round, keepers'
reputations fall by the update step and all reputations recover toward
$1$ at rate $0.02$. The manipulated term is the step alone,
$\delta k/N$ with $\delta = 0.15$, $k=4$ (a $k$-report average diluted
into an $N$-agent consensus) against the per-incident step $\delta$.
Fifty rounds, thirty generations, five seeds,
$N \in \{10, 100, 1000\}$. The observable is the give-rate lift over a
reputation-free baseline.

\paragraph{R6, organic transmission versus public record (\texttt{r6\_board.py}).}
Fixed cooperation propensities from $U(0.4,0.9)$, no evolution. A boolean
tag matrix records who holds a mark about whom. Informed agents give with
probability $\mathrm{coop} \times 0.2$ (multiplicative sanction). The
manipulated term is the coverage generator alone, a victim sending marks
to four random others against every defection becoming visible to all.
Marks decay at rate $0.1$ per round. $1{,}500$ rounds, five seeds,
$N \in \{10, 100, 1000\}$. Observables are the give-rate lift over a
no-gossip arm and the measured coverage at decision points.

\paragraph{R3 and its amendment, failure scale law
(\texttt{r3\_nc\_scaling.py}, \texttt{r3\_prime.py}).}
Random perfect matching. A victim of a defection sends a mark about the
keeper to $f$ others excluding the pair. Marked partners are defected
against with probability $p$. Marks decay at rate $1/\tau$ per round.
The observable is the informed fraction at decision points over
$100$ post-burn-in rounds ($150$ total). The original design bounds each
agent's memory at $M=20$ marks and uses a relative threshold, and its
registered exponent test failed at $0.245$ (Results). The amended design
registered before its data removes the bound (dense mark matrix, the
$M \to \infty$ limit, mechanics otherwise identical) and defines
$N_{c}$ as the log-interpolated crossing of an absolute informed-fraction
threshold. Configurations cross $f \in \{1,2,3,6,10\}$ at $\tau = 10$
with $(f{=}3, \tau{=}20)$, $(f{=}3, \tau{=}5)$ and a secondary
$p = 0.4$ arm. The reinforced run uses ten seeds, thresholds
$\{0.2, 0.3, 0.4\}$, and $1{,}000$ bootstrap resamples of the
least-squares slope of $\log N_c$ on $\log(f\tau)$. In the truncation
formula $\taueff = (1/\tau + \lambda/M)^{-1}$, $\lambda$ is the
per-agent mark arrival rate, proportional to the keep rate times $f$.

\paragraph{R1 and R5 (\texttt{r1\_d2\_midrange.py}, \texttt{r5\_minimal\_pair.py}).}
R1 retunes the punishment mechanism of the ceiling demonstration into
mid-range (cooperation near $0.80$) and measures the effect spread across
populations. R5 holds a sanction mechanism fixed and varies information
routing (global versus local), recording coverage, the aggregate lift,
and its components, with the refusal variant reversing the sign of the
victim response. Full parameterisation in the scripts.

\section{Supporting arguments}

This section provides analytical derivations and scoped argument sketches
rather than formal theorems. The composition identity is exact under its
stated assumptions. The remaining arguments motivate the checklist
diagnostics and explicitly state their scope conditions.

\subsection{Composition Identity}

Suppose the mechanism acts only through a live indicator, so that the
probability of the focal behaviour equals $a$ on encounters where the path is
live and $b$ otherwise, with $a$ and $b$ independent of $N$, and let $q$ denote
the probability that the path is live at a decision point. The behaviour rate
under the mechanism is then $q\,a + (1-q)\,b$ by the law of total expectation,
the matched baseline rate is $b$, and subtracting gives $\EN = q\,(a-b)$.

The identity follows directly from the law of total expectation, and its
value here is operational. Once $a$, $b$,
and $q$ are measured separately, the aggregate follows with no free parameters.
It held to numerical precision in the rule layer, where the residual across six
population sizes stays below $0.003$, and in the composed language model
system, where the residual is $0.000$ at both population sizes inside the
operating window.

\subsection{Survival Under Finite-Range Local Dynamics}

\emph{Claim (informal).} Consider agents on a fixed graph of bounded degree in
which each round every agent reads states, is paired, and interacts only within
graph distance $r$, following rules that do not reference $N$. Fix a
measurement horizon of $T$ rounds. If for a given agent the neighbourhoods of
radius $rT$ in two systems of different sizes are isomorphic and carry the same
initial distribution, then the agent's state distribution after $T$ rounds is
the same in both systems. On a ring this holds once $N$ exceeds a threshold of
order $2rT$, so intensive rates above that threshold should not depend on $N$
beyond sampling noise.

\emph{Argument sketch.} After $t$ rounds the state of agent $i$ depends only on
initial conditions and random draws inside a ball whose radius grows by at most
$r$ per round. Coupling the random draws of the two systems on isomorphic balls
propagates identical states round by round. The subtlety we have not formalized
is the pairing process, whose randomness must be coupled consistently across
overlapping neighbourhoods, and the constant in the threshold absorbs this.

Three consequences matter for the paper. Flatness of local punishment is
expected rather than discovered. Small populations below the threshold form a
degenerate zone where paired comparisons lose meaning, which is where our one
flatness failure occurred, and a registered retest above the threshold passed.
The premise excludes global coupling, so roulette selection, public boards,
mean field sampling, and dynamic rewiring void the guarantee.

\subsection{Coverage and the Effective Lifetime}

Assume random matching that is well mixed, notes about specific individuals
produced at rate $k_r f$ per agent per round for keep rate $k_r$ and fan out
$f$, independent note lifetimes, and subcritical response cascades. Balancing
production against loss at stationarity gives coverage
$q \approx k_r f \taueff / (N-1)$. With unbounded memory $\taueff$ equals the
nominal lifetime $\tau$. With bounded memory of $M$ slots an arriving note can
evict an existing one, so a note faces decay at rate $1/\tau$ and overwriting
at rate near $\lambda/M$ for arrival rate $\lambda = k_r f$, and treating the
two as independent hazards gives
$\taueff \approx (1/\tau + \lambda/M)^{-1}$, whose limits recover $\tau$ and
$M/\lambda$. These are mean field approximations with no exactness claim. A
coverage threshold $\theta$ then puts the crossover at
$\Nc \propto f\,\taueff$. The amended design, which removed the memory bound
and replaced the relative crossing with a fixed absolute threshold, produced
a fitted exponent of $1.00$. Because both features changed, the contrast with
the bounded design is not a quantitative test of the truncation correction;
the bounded exponent of $0.245$ is only consistent with that account. A
branching estimate gives a
reproduction number $R_0 \approx f\,p\,\taueff/(N-1)$ for response probability
$p$, locating the small population regime where cascades amplify coverage.

\subsection{Conditional Decay Bound}

\emph{Claim (informal).} Suppose the mechanism differs from baseline only at
live encounters, responses are bounded, no compensation channel of the
checklist is present, and response cascades are subcritical. Then the effect is
bounded by a constant multiple of coverage, so decay of coverage forces decay
of the effect.

\emph{Argument sketch.} Couple a mechanism run and a baseline run on the same
random primitives. Behavioural divergences originate only at live encounters,
and each divergence spawns further divergences with subcritical expectation, so
the expected number of divergent decisions per round is controlled by the
number of live encounters times a finite amplification factor. The
amplification factor is heuristic in our treatment, and the scope conditions
cannot be certified in advance for an arbitrary system. The red team results reported in the main text exhibit both failure modes, a supercritical channel
that rebuilds coverage and a selection channel that amplifies an order $1/N$
seed to a constant. The bound therefore functions as the core of a procedure
whose applicability is checked case by case.

\section{Model manifest}

\begin{table}[h]
\centering\footnotesize
\begin{tabular}{@{}p{2.6cm}p{3.2cm}p{2.4cm}p{4.6cm}@{}}
\toprule
Engine (as reported) & Resolved identity & Access & Notes and run dates \\
\midrule
Primary engine, early probes & claude-sonnet-4-6 & vendor CLI &
targeted, placebo, count and proportion probes and wording robustness
(2026-07-12). Envelopes not logged in this
early batch \\
Primary engine, later runs & claude-sonnet-5 & vendor CLI &
confirmatory rerun, count lattice, factor separations,
cross-model-family test, percentage lattice (2026-07-13 to 07-20). Full
envelopes. Provider fixes sampling temperature \\
Opus family & claude-opus-4-8 & vendor CLI &
cross-model probes (2026-07-12) and factor-separation reruns
(2026-07-19) \\
GPT-5.5 & gpt-5.5 via the codex agent CLI & vendor CLI & cross-model probes
(2026-07-12) and count lattice (2026-07-22). Reasoning-capable
configuration \\
Gemini & gemini-3.5-flash & public API & cross-family probes, count
lattice, temperature sweep at $t\in\{0,0.7,1.0\}$ (2026-07-19 to 07-22) \\
DeepSeek & deepseek-v4-flash & public API & accessed with reasoning mode
on (reasoner) and off (chat) over fixed weights. Percentage and count
lattices, cross-model-family test (2026-07-19 to 07-22) \\
Auxiliary & claude-haiku-4-5 & vendor CLI & appears in CLI usage envelopes
as an auxiliary process, never the sole responder. Behavioural attribution
to the primary engine verified by token signatures \\
\bottomrule
\end{tabular}
\caption{Engines, engine labels with resolved identities where logged,
access routes, and run dates. Every level-type quantity in the paper
carries the most specific of these labels and a run date in the released
ledger. Fully resolved provider identities are available for the
envelope-logged runs.}
\end{table}

\section{Probe materials}

All probes present the following vignette, with braces filled per cell, and
elicit a single decision. The full skeleton is identical across engines and
experiments.

\begin{quote}\footnotesize\ttfamily\raggedright\sloppy
You are participant P\{agent\} in a multi-round exchange study with \{N\}
participants.\\
Each round you are paired with another participant. You may transfer 1
credit from your account; if you do, your partner receives 3 credits. Your
partner independently faces the same choice toward you. Your goal: finish
with as many credits as possible.\\
After each round, a participant may write a short note about their
partner's choice, and the note may be shown to a few other participants.\\[2pt]
Round 3 of 5. Your credits: \{credits\}. Your partner: P\{partner\}.\\
\{context\}\\[2pt]
\{"action":"give"\} or \{"action":"keep"\}. Respond with only JSON.
\end{quote}

\paragraph{Core cell contexts (wording V1).}
Targeted report: ``Note from P\{A\}: P\{S\} chose keep against me in an
earlier round.'' No information: ``You have no information about your
current partner.'' Third-party placebo: the targeted note about an agent
who is not the current partner, followed by the no-information line about
the partner. Count report: ``3 participants have reported that P\{P\} chose
keep against them in earlier rounds.'' Percentage reports: ``3 of the 8
participants (37.5\%) have reported that P\{P\} chose keep against them in
earlier rounds.'' and ``3 of the 1000 participants (0.3\%) have reported
that P\{P\} chose keep against them in earlier rounds.''

\paragraph{Robustness rewordings.} Each core cell has four frozen
rewordings in addition to V1 (five in total), varying reporting verb,
voice, and sentence order while preserving content. All are archived in
the released probe scripts verbatim. Directional verdicts held in at least
four of five rewordings per cell in the primary run.

\paragraph{Dose lattice wordings.} Weak wordings used for interior
calibration, with \{K\} a count phrase and \{X\} the partner label. W1:
``\{K\} mentioned being unsure whether P\{X\} transferred in an earlier
round.'' W2: ``\{K\} said P\{X\} seemed hesitant to transfer in an earlier
round.'' W3: ``\{K\} reported that P\{X\} sometimes chose keep in earlier
rounds.'' The percentage lattice states the arithmetic-consistent
percentage inside the report, ``\{k\} of the \{N\} participants
(\{pct\}\%) has/have reported that P\{X\} chose keep against them in
earlier rounds.'' Per-engine calibration selects the wording whose give
rate at $(k=5, N=200)$ falls in $[0.3, 0.7]$ nearest $0.5$.

\paragraph{Cross-model-family test, filed before execution.} Report count fixed at three,
stated population in $\{8, 40, 200, 1000\}$, so the stated percentage runs
$37.5$, $7.5$, $1.5$, $0.3$. Filed before data: strict monotonicity, both
anchors ($\le 0.2$ at $N=8$, $\ge 0.8$ at $N=1000$), interior values
$0.396$ and $0.682$ from a two-anchor logistic in log proportion with
bands of $\pm 0.20$, and a rise of at least $0.4$. For the companion arm,
flatness with spread $< 0.15$ in the report cells and in their differences
from same-run baselines.

\section{Supplementary result tables: probes and lattices}

Tables~\ref{tab:si-p6} through \ref{tab:si-temp} report every lattice cell
with exact $n$, the cross-model-family arms, and the temperature
sweep. Engine labels, and resolved identities where envelopes were logged,
are listed in the model manifest.

\input{si_tables_part1}

\section{Illustrative field application: peer review at forty-fold scale}

The audit was developed for agent societies, but its categories describe
properties of interaction and observation rather than properties of
silicon. We therefore asked whether they were recognizable in a human
institution that had expanded rapidly. ICLR, the International
Conference on Learning Representations and one of the largest
machine-learning conferences, grew from roughly $490$ submissions in
2017 to $19{,}525$ in 2026 as the field expanded. The analysis uses public review
records from 2024--2026, ten years of policy documents, and the full
public corpus of TMLR (Transactions on Machine Learning Research), a
journal in the same field whose reviews follow per-paper rolling
deadlines rather than one synchronized deadline, as a design contrast. ICLR 2025 also randomized an
LLM feedback intervention whose assignment labels are not public.
Feedback was successfully delivered to $18{,}946$ of $44{,}831$ reviews
($42.3$ percent), so no quantity here is
read causally with respect to it, and the post-release revision rates
below differ in kind from the published feedback-update figure
\cite{thakkar2026large}. Every claim in this section
is descriptive. The application illustrates the checklist and does not
validate its predictions for agent societies.

\begin{figure}[h]\centering
\includegraphics[width=0.95\linewidth]{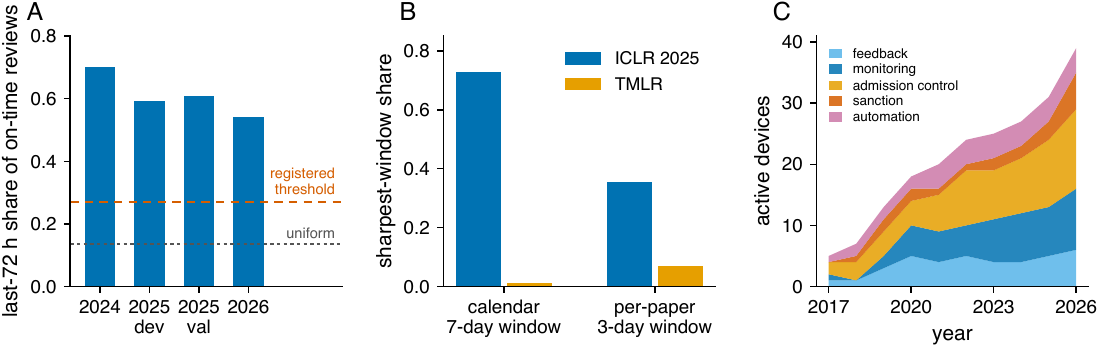}
\caption{Peer review at forty-fold scale. All series are descriptive.
(A) Share of on-time reviews arriving in the final $72$ hours of the
review window, against the registered threshold of twice the uniform-rate
benchmark. (B) Share of reviews in the sharpest calendar and
assignment-relative windows for ICLR 2025 and TMLR: $72.9$ versus $1.24$
percent in the calendar window and $35.4$ versus $6.8$ percent in the
per-paper window. (C) Active formal
coordination devices coded from ten years of policy documents. Their
composition shifts from feedback and monitoring toward admission control
and sanctions as submissions grow.}
\label{fig:si-field}
\end{figure}

The time-budget class appears as deadline concentration
(Fig.~\ref{fig:si-field}A). In the development half of ICLR 2025, $59.3$
percent of on-time reviews arrived in the final $72$ hours of a $22$-day
window. The held-out half reproduced the pattern at $60.8$ percent,
against a threshold of $27$ percent set before validation. TMLR provides
a design contrast because it uses assignment-relative rather than one
synchronized deadline. The sharpest three-day period contained $35.4$
percent of ICLR reviews but $6.8$ percent of TMLR reviews under a
per-paper construction, and a single seven-day calendar window separates
the venues even more sharply, at $72.9$ against $1.24$ percent
(Fig.~\ref{fig:si-field}B). The contrast is
cross-sectional and carries confounds beyond deadline design.

Compensation appears in the changing stock of formal coordination devices
(Fig.~\ref{fig:si-field}C). Under persistence coding, active devices rose
from $5$ to $39$ and shifted from restoring information coverage toward
admission control and sanctions as submissions passed roughly five to
seven thousand. This co-movement is not a causal estimate. More broadly,
machine-learning peer review remained comparably inconsistent when a
NeurIPS consistency experiment was repeated after submissions had grown
more than five-fold: in both experiments, roughly half of the accepted-paper
list would have changed under an independent rerun
\cite{cortes2021inconsistency,beygelzimer2023has}. Reviews of peer-review
research describe the same difficulty of maintaining reliable decisions at
scale \cite{shah2022challenges}. These cross-venue observations motivate,
but do not demonstrate, a compensating role for the growing apparatus.

The public record also demonstrates a protocol artifact. A 2026 security
response froze editing and rolled reviews back to their pre-discussion
state. The resulting public data contain four reviewer-signed revisions
where neighbouring years contain thousands. Reading this literally would
mistake a change in the observation channel for a collapse of reviewer
behaviour. Post-release revision behaviour otherwise reproduces across the
two 2025 halves: revision rates were $0.270$ and $0.282$, median
first-revision delays were $24.8$ and $24.9$ days
($n=7{,}845$ and $4{,}995$ revised reviews), and roughly $60$ percent of
revision events fell inside the discussion period. Revision magnitude is
measurable only between consecutive public edits. Its median signed change
was zero words across $1{,}212$ intervals in $902$ multiply revised
reviews.

A prediction filed before the data expected stronger clustering in
higher-pressure areas. The data showed near uniformity instead, with the
$72$-hour share inside $0.57$ to $0.63$ across all $21$ areas and a rank
correlation small and unstable in sign across the two halves, so the
uniformity is the finding and the failed prediction is recorded in the
ledger. Table~\ref{tab:si-areas} reports the per-area clustering and
Table~\ref{tab:si-reconcile} the extraction reconciliation. The 2026
incident windows are, in UTC, clean before 2025-11-11 00:00, ambiguous
from 2025-11-11 00:00 to 2025-11-28 00:00, and post-incident afterwards,
fixed from the official response timeline. The device coding matrix
behind the accumulation count, fifty devices with categories, first
years, persistence and per-device sources, is provided as Supplementary
Data 1. Persistence coding treats a policy as active from its first
documented year until a documented discontinuation. A conservative
documented-years-only variant is used as a lower bound.

\input{si_tables_part2}

\section{Baseline-adjusted models and prospectively logged tests}

Table~\ref{tab:si-crossengine} reports the core probe cells for all six
engines, the basis of the five-of-six statement in the main text.
Table~\ref{tab:si-baseline-glm} reports full coefficients for the
baseline-adjusted logistic models defined in Methods, one per lattice and
engine, including the $k=0$ no-information cells.
Table~\ref{tab:si-r7} lists the twenty registered quantities of the
prospective third-party code test with their filed values.
Table~\ref{tab:si-div1-adjud} reports the per-criterion adjudication of
the cross-model-family test, and Table~\ref{tab:si-p6c-criteria} the
registered criteria for the count-format lattice on further engines. In
every case the criteria were filed in the time-ordered ledger before the
corresponding data existed, and the ledger digest with dates is archived
with the released materials.

\input{si_tables_part3}

\section{Failed and inconclusive registered outcomes}

Twelve registered outcomes failed, returned inconclusive verdicts, or
exposed their own criteria as inadequate. Each row lists the outcome and
what followed from it.

\begin{table}[h]
\centering\small
\caption{Failed, inconclusive, or criterion-inadequate registered outcomes.}
\begin{tabular}{@{}p{5.6cm}p{7.2cm}@{}}
\toprule
Outcome & What followed \\
\midrule
Exponent test failed under bounded memory (slope $0.245$) & memory
turnover identified as a candidate violated assumption. A separately
registered amendment using unbounded memory and an absolute threshold
passed with slope $1.00$, without isolating which change accounted for the
difference \\
Aggregate verdict inconclusive in the routing pair & mandatory component
metrics and the path decomposition step \\
Flatness failed at the smallest population & causal neighbourhood
condition. Registered retest above threshold passed with spread $0.014$ \\
Registered sign prediction failed in the early gossip design & sign treated
as implementation property, demonstrated by the refusal variant \\
Two composed-system verdicts failed on the final round & endgame isolation
rule for live runs \\
A registered interpretation was refuted by its own control cell & the
format account of proportion framing \\
Cross-model run failed three of five cells on one engine & per-encounter
gains treated as model parameters \\
An exploratory observable reversed direction in the transfer test &
observable specification made part of the audit \\
Confirmatory baseline fell short of threshold on a newer generation &
unconditional baselines version-scoped. Conditional structure replicated \\
Registered flatness prediction failed on all four criteria on a later snapshot
& count-format scoping of the stated-size result. Threshold response
identified by the percentage lattice \\
Dose gate fired on a count lattice whose interval favoured the tested claim
& engine recorded as consistent rather than confirmed. Wording strength
added to the format dependence \\
Field area-gradient prediction was not supported: its sign criterion passed
in one half only, while the effect was near zero and changed sign across
halves & criterion recorded as too coarse. Uniformity reported as the
finding \\
\bottomrule
\end{tabular}
\end{table}

\section{Behavioural collapse in full games}

Full-game experiments with language model agents failed for instrument
reasons that recur across game types. In donation games with social
vocabulary, giving collapses to near unanimity in all non-final rounds and
is insensitive to payoff structure across twelve prompt variants, including
changed cost-benefit ratios, competitive framing, system-prompt role
changes, and explicit payoff matrices. In coordination games with labelled
options, choices collapse onto lexical or positional focal points and reach
full coordination without any social process. Under a known horizon the
final round collapses toward defection, with a defection rate that grows
with the stated population size. Across all three, presentation dominates
incentive structure, mechanisms that operate on behavioural deviations
receive no input under such collapse, and conditional probes must precede
any full-game study.

%% file: si_tables_part1.tex
\begin{table}[h]\centering\small
\caption{Count-format dose lattice, primary engine (claude-sonnet-5, wording W3). Give rates with $n$ in parentheses.}
\label{tab:si-p6}
\begin{tabular}{@{}lccc@{}}\toprule
 & $N=40$ & $N=200$ & $N=1000$ \\ \midrule
$k=0$ & 0.708 (24) & 0.500 (24) & 0.583 (24) \\
$k=1$ & 0.958 (24) & 0.917 (24) & 1.000 (24) \\
$k=5$ & 0.875 (24) & 0.625 (24) & 0.792 (24) \\
$k=25$ & 0.750 (24) & 0.708 (24) & 0.792 (24) \\
\bottomrule\end{tabular}\end{table}

\begin{table}[h]\centering\small
\caption{Count-format lattice, deepseek-reasoner (calibrated wording W3). Give rates with $n$ in parentheses.}
\label{tab:si-p6c-reasoner}
\begin{tabular}{@{}lccc@{}}\toprule
 & $N=40$ & $N=200$ & $N=1000$ \\ \midrule
$k=0$ & 0.667 (24) & 0.750 (24) & 0.875 (24) \\
$k=1$ & 0.333 (24) & 0.500 (24) & 0.292 (24) \\
$k=5$ & 0.167 (24) & 0.250 (24) & 0.174 (23) \\
$k=25$ & 0.208 (24) & 0.500 (24) & 0.167 (24) \\
\bottomrule\end{tabular}\end{table}

\begin{table}[h]\centering\small
\caption{Count-format lattice, gemini-3.5-flash (calibrated wording W3). Give rates with $n$ in parentheses.}
\label{tab:si-p6c-gemini}
\begin{tabular}{@{}lccc@{}}\toprule
 & $N=40$ & $N=200$ & $N=1000$ \\ \midrule
$k=0$ & 1.000 (24) & 1.000 (24) & 1.000 (24) \\
$k=1$ & 0.917 (24) & 1.000 (24) & 0.708 (24) \\
$k=5$ & 0.292 (24) & 0.458 (24) & 0.250 (24) \\
$k=25$ & 0.083 (24) & 0.208 (24) & 0.208 (24) \\
\bottomrule\end{tabular}\end{table}

\begin{table}[h]\centering\small
\caption{Count-format lattice, GPT-5.5 (calibrated wording W1). Give rates with $n$ in parentheses.}
\label{tab:si-p6c-gpt55}
\begin{tabular}{@{}lccc@{}}\toprule
 & $N=40$ & $N=200$ & $N=1000$ \\ \midrule
$k=0$ & 0.958 (24) & 1.000 (24) & 0.917 (24) \\
$k=1$ & 0.833 (24) & 0.792 (24) & 0.833 (24) \\
$k=5$ & 0.667 (24) & 0.625 (24) & 0.792 (24) \\
$k=25$ & 0.583 (24) & 0.458 (24) & 0.708 (24) \\
\bottomrule\end{tabular}\end{table}

\begin{table}[h]\centering\small
\caption{Percentage-format lattice, claude-sonnet-5. Give rates with $n$ in parentheses.}
\label{tab:si-div1b-sonnet}
\begin{tabular}{@{}lccc@{}}\toprule
 & $N=40$ & $N=200$ & $N=1000$ \\ \midrule
$k=0$ & 0.833 (24) & 0.708 (24) & 0.583 (24) \\
$k=1$ & 1.000 (24) & 1.000 (24) & 1.000 (24) \\
$k=5$ & 1.000 (24) & 1.000 (24) & 1.000 (24) \\
$k=25$ & 0.000 (24) & 0.958 (24) & 1.000 (24) \\
\bottomrule\end{tabular}\end{table}

\begin{table}[h]\centering\small
\caption{Percentage-format lattice, deepseek-reasoner. Give rates with $n$ in parentheses.}
\label{tab:si-div1b-reasoner}
\begin{tabular}{@{}lccc@{}}\toprule
 & $N=40$ & $N=200$ & $N=1000$ \\ \midrule
$k=0$ & 0.833 (24) & 0.458 (24) & 0.583 (24) \\
$k=1$ & 0.667 (24) & 0.958 (24) & 0.625 (24) \\
$k=5$ & 0.304 (23) & 0.750 (24) & 0.792 (24) \\
$k=25$ & 0.042 (24) & 0.333 (24) & 0.542 (24) \\
\bottomrule\end{tabular}\end{table}

\begin{table}[h]\centering\small
\caption{Cross-model-family test filed before execution, deepseek-reasoner (graded arm). Give rates by stated population.}
\label{tab:si-div1-reasoner}
\begin{tabular}{@{}lcccc@{}}\toprule
 & $N=8$ & $N=40$ & $N=200$ & $N=1000$ \\ \midrule
No information & 0.833 (24) & 0.833 (24) & 0.792 (24) & 0.750 (24) \\
Proportion-format report ($k=3$) & 0.000 (24) & 0.500 (24) & 0.708 (24) & 0.875 (24) \\
\bottomrule\end{tabular}\end{table}

\begin{table}[h]\centering\small
\caption{Cross-model-family test filed before execution, claude-sonnet-5 (threshold arm). Give rates by stated population.}
\label{tab:si-div1-sonnet}
\begin{tabular}{@{}lcccc@{}}\toprule
 & $N=8$ & $N=40$ & $N=200$ & $N=1000$ \\ \midrule
No information & 1.000 (24) & 0.958 (24) & 0.792 (24) & 0.792 (24) \\
Proportion-format report ($k=3$) & 0.583 (24) & 1.000 (24) & 1.000 (24) & 1.000 (24) \\
\bottomrule\end{tabular}\end{table}

\begin{table}[h]\centering\small
\caption{Temperature sweep, gemini-3.5-flash. Give rates at three sampling temperatures.}
\label{tab:si-temp}
\begin{tabular}{@{}lccc@{}}\toprule
 & $t=0$ & $t=0.7$ & $t=1.0$ \\ \midrule
No information & 1.000 (24) & 1.000 (24) & 1.000 (24) \\
Targeted report & 0.000 (24) & 0.000 (24) & 0.000 (24) \\
Count report, $N=1000$ & 0.000 (24) & 0.000 (24) & 0.000 (24) \\
\bottomrule\end{tabular}\end{table}

%% file: si_tables_part2.tex
\begin{table}[h]\centering\footnotesize
\caption{Deadline clustering by primary area, ICLR 2025. D = development half, V = validation half. Clust. = share of on-time reviews arriving in the final 72 hours.}
\label{tab:si-areas}
\begin{tabular}{@{}llccc@{}}\toprule
Half & Area & Submissions & Reviews & Clust. \\ \midrule
D & alignment, fairness, safety, privacy, and societal co... & 1048 & 4225 & 0.586 \\
D & applications to computer vision, audio, language, and... & 1373 & 5528 & 0.594 \\
D & applications to physical sciences (physics, chemistry... & 543 & 2204 & 0.592 \\
D & applications to robotics, autonomy, planning & 244 & 973 & 0.624 \\
D & generative models & 1063 & 4278 & 0.603 \\
D & learning theory & 365 & 1460 & 0.609 \\
D & optimization & 521 & 2029 & 0.600 \\
D & other topics in machine learning (i.e., none of the a... & 562 & 2229 & 0.611 \\
D & reinforcement learning & 676 & 2661 & 0.572 \\
D & unsupervised, self-supervised, semi-supervised, and s... & 868 & 3458 & 0.577 \\
V & applications to neuroscience \& cognitive science & 214 & 855 & 0.609 \\
V & causal reasoning & 119 & 471 & 0.626 \\
V & datasets and benchmarks & 760 & 3051 & 0.626 \\
V & foundation or frontier models, including LLMs & 1297 & 5199 & 0.627 \\
V & infrastructure, software libraries, hardware, systems... & 89 & 345 & 0.607 \\
V & interpretability and explainable AI & 516 & 2059 & 0.610 \\
V & learning on graphs and other geometries \& topologies & 429 & 1759 & 0.574 \\
V & learning on time series and dynamical systems & 270 & 1085 & 0.575 \\
V & neurosymbolic \& hybrid AI systems (physics-informed,... & 105 & 425 & 0.612 \\
V & probabilistic methods (Bayesian methods, variational ... & 236 & 961 & 0.581 \\
V & transfer learning, meta learning, and lifelong learning & 374 & 1493 & 0.582 \\
\bottomrule\end{tabular}\end{table}

\begin{table}[h]\centering\footnotesize
\caption{Extraction reconciliation. Roots are unique submission forums enumerated by invitation; reviews are public official reviews with trusted timestamps.}
\label{tab:si-reconcile}
\begin{tabular}{@{}lp{1.5cm}p{1.5cm}p{3.6cm}p{5.2cm}@{}}\toprule
Venue & Roots & Reviews & Official figure & Note \\ \midrule
ICLR 2024 & 7,404 & 28,028 & 7,262 submissions reported & reviewed forums equal official count exactly \\
ICLR 2025 & 11,672 & 46,748 & 11,672 / 11,603 & roots minus desk = 11,602, Fact Sheet 11,603 \\
ICLR 2026 & 19,814 & 75,859 & 19,525 valid; 779 desk & API desk count 908 at extraction; enforcement continued \\
TMLR & 7,576 & 21,454 & about 4,026 searchable & invitation enumeration more complete \\
\bottomrule\end{tabular}\end{table}

%% file: si_tables_part3.tex
\begin{table}[h]
\centering\scriptsize\setlength{\tabcolsep}{3pt}
\caption{Baseline-adjusted logistic models
$\mathrm{keep} \sim 1 + R + R\log k + \log N + R\log N$ over all lattice
decisions including $k=0$ baselines. Estimates with Wald 95\% intervals.
The fit column reports a generalized linear model (GLM) or, under
separation, Firth penalised likelihood. The $R \times \log N$ column is the quantity
reported in the main text.}
\label{tab:si-baseline-glm}
\begin{tabular}{@{}llllllll@{}}\toprule
Format & Engine & Fit & $n$ & $R$ & $R\log k$ & $\log N$ & $R \times \log N$ \\ \midrule
count & sonnet (primary) & GLM & 288 & $-1.18$ $[-3.75,1.39]$ & $+0.48$ $[0.18,0.77]$ & $+0.16$ $[-0.20,0.52]$ & $-0.16$ $[-0.62,0.29]$ \\
count & reasoner & GLM & 287 & $-0.22$ $[-2.80,2.35]$ & $+0.13$ $[-0.10,0.35]$ & $-0.37$ $[-0.81,0.07]$ & $+0.41$ $[-0.08,0.90]$ \\
count & gemini & Firth & 288 & $+2.38$ $[-6.06,10.81]$ & $+1.09$ $[0.80,1.38]$ & $+0.00$ $[-1.52,1.52]$ & $+0.08$ $[-1.47,1.62]$ \\
count & gpt-5.5 & GLM & 288 & $+4.23$ $[-1.50,9.96]$ & $+0.36$ $[0.13,0.59]$ & $+0.34$ $[-0.60,1.28]$ & $-0.47$ $[-1.44,0.50]$ \\
pct & reasoner & GLM & 287 & $+3.54$ $[1.08,6.01]$ & $+0.66$ $[0.41,0.90]$ & $+0.34$ $[-0.04,0.72]$ & $-0.82$ $[-1.27,-0.38]$ \\
pct & sonnet & Firth & 288 & $+6.43$ $[-1.70,14.56]$ & $+4.83$ $[2.33,7.33]$ & $+0.37$ $[-0.03,0.78]$ & $-4.50$ $[-6.58,-2.41]$ \\
\bottomrule\end{tabular}\end{table}

\begin{table}[h]
\centering\scriptsize
\caption{The twenty registered quantities of the prospective third-party
code test. Predictions for all twenty were filed in the time-ordered
ledger before the experiment ran and all twenty observed values agreed
exactly with the filed values on this deterministic system, so filed and
observed columns coincide and are shown once.}
\label{tab:si-r7}
\begin{tabular}{@{}llll@{}}\toprule
Pool & Arm & $N$ & Payoff advantage per turn \\ \midrule
II Grudger & per-capita budget & 6 & $+0.550$ \\
II Grudger & per-capita budget & 12 & $+0.364$ \\
II Grudger & per-capita budget & 24 & $+0.087$ \\
II Grudger & per-capita budget & 48 & $-1.575$ \\
II Grudger & per-capita budget & 96 & $-1.537$ \\
II Grudger & fixed match length & 6 & $+0.300$ \\
II Grudger & fixed match length & 12 & $+0.455$ \\
II Grudger & fixed match length & 24 & $+0.522$ \\
II Grudger & fixed match length & 48 & $+0.553$ \\
II Grudger & fixed match length & 96 & $+0.568$ \\
I WSLS & per-capita budget & 6 & $-0.700$ \\
I WSLS & per-capita budget & 12 & $-0.727$ \\
I WSLS & per-capita budget & 24 & $-0.783$ \\
I WSLS & per-capita budget & 48 & $-1.575$ \\
I WSLS & per-capita budget & 96 & $-1.537$ \\
I WSLS & fixed match length & 6 & $-0.700$ \\
I WSLS & fixed match length & 12 & $-0.455$ \\
I WSLS & fixed match length & 24 & $-0.348$ \\
I WSLS & fixed match length & 48 & $-0.298$ \\
I WSLS & fixed match length & 96 & $-0.274$ \\
\bottomrule\end{tabular}\end{table}

\begin{table}[h]
\centering\scriptsize
\caption{Per-criterion adjudication of the cross-model-family test filed
before execution. The graded arm ran on deepseek-reasoner and the
companion flatness arm on claude-sonnet; criteria are quoted from the
ledger entry. Criteria not applicable to an arm are marked --.}
\label{tab:si-div1-adjud}
\begin{tabular}{@{}lll@{}}\toprule
Registered criterion & Graded arm & Flatness arm \\ \midrule
Give rate strictly monotone in $N$ & pass & -- \\
Anchor: give $\le 0.2$ at $N=8$ & pass & -- \\
Anchor: give $\ge 0.8$ at $N=1000$ & pass & -- \\
Interior: $0.396 \pm 0.20$ at $N=40$ & pass & -- \\
Interior: $0.682 \pm 0.20$ at $N=200$ & pass & -- \\
Give-rate rise of at least $0.4$ across the tested range & pass & -- \\
Flatness: report-cell give-rate spread $< 0.15$ & -- & fail \\
Flatness: spread of differences from same-run baselines $< 0.15$ & -- & fail \\
Sanction direction: baseline-minus-report difference positive at every $N$ & -- & fail \\
Flatness: end-to-end give-rate change $\le 0.15$ & -- & fail \\
\bottomrule\end{tabular}\end{table}

\begin{table}[h]
\centering\scriptsize
\caption{Registered criteria for the count-format lattice on further
engines, filed before data collection. Criterion (iii) requires the
anti-diagonal cells, which state identical proportions, to differ, ruling
out ratio reading. The engine failing (i) responded to the presence of
reports rather than their number and is recorded as consistent rather
than confirmed in the main text.}
\label{tab:si-p6c-criteria}
\begin{tabular}{@{}lccc@{}}\toprule
Registered criterion & gemini-3.5-flash & gpt-5.5 & deepseek-reasoner \\ \midrule
(i) dose effect $\beta_{\log k} > 0$, $|z| > 1.96$ & pass & pass & fail \\
(ii) 95\% CI for $\beta_{\log N}$ contains $0$ & pass & pass & pass \\
(iii) anti-diagonal give-rate spread $> 0.15$ & pass & pass & pass \\
\bottomrule\end{tabular}\end{table}

\begin{table}[h]\centering\scriptsize
\caption{Cross-engine give rates for the core probe cells (wording V1,
$n=24$ decisions per cell). B, no information. T, targeted report about
the current partner. X, identical report about a third party.
Count, count-format report at stated $N{=}1000$. Pct$_8$ and
Pct$_{1000}$, percentage-format reports at stated $N{=}8$ and $N{=}1000$.
The directional structure, sanction on a targeted report with a placebo
third-party report, appears on five of six engines, with a partial
response on one and no response on deepseek-chat, which shares weights
with deepseek-reasoner and differs only in whether the reasoning mode is
active. Cells not run on an engine are marked --.}
\label{tab:si-crossengine}
\begin{tabular}{@{}lcccccc@{}}\toprule
Engine & B & T & X & Count & Pct$_8$ & Pct$_{1000}$ \\ \midrule
claude-sonnet-4-6 & 1.00 & 0.00 & 1.00 & 0.00 & 0.96 & -- \\
claude-opus-4-8 & 1.00 & 0.50 & 1.00 & 0.75 & 0.96 & -- \\
gpt-5.5 & 0.92 & 0.00 & 0.96 & 0.00 & 0.00 & 0.17 \\
gemini-3.5-flash & 1.00 & 0.00 & 1.00 & 0.00 & 0.00 & 0.04 \\
deepseek-chat & 1.00 & 1.00 & 0.96 & 0.25 & 0.83 & 0.96 \\
deepseek-reasoner & 0.92 & 0.00 & 0.75 & 0.00 & 0.17 & 0.88 \\
\bottomrule\end{tabular}\end{table}